\documentclass[twocolumn,showpacs,amsmath,amssymb]{revtex4-1}
\input{epsf}

\usepackage{graphicx}
\usepackage{dcolumn}
\usepackage{array}                      
\usepackage{bigstrut}
\usepackage{longtable}
\usepackage{rotating,booktabs}
\usepackage{booktabs,threeparttable}
\usepackage{color}
\usepackage{bm}
\def\bdot{\mbox{\boldmath{$\cdot$}}}
\usepackage{color}
\definecolor{MyDarkBlue}{rgb}{0,0.08,0.55}

\def\bp{{\bf p}}
\def\br{{\bf r}}
\begin{document}

\title{Theory for the Rydberg states of helium: quantum defect extensions and comparison with experiment up to $n = 102$ for the singlet and triplet $P$-states}

\author{G. W. F. Drake$^{1,*}$}
\author{Aaron T. Bondy$^2$}

\affiliation { $^1$ Department of Physics, University of Windsor, Windsor, Ontario, Canada N9B 3P4}
\affiliation{ $^2$ Department of Physics and Astronomy, Drake University, Des Moines, Iowa 50311}

\begin{abstract}

High precision variational calculations for helium in Hylleraas coordinates are used to obtain a combination of quantum defect expansions for the nonrelativistic energy and $1/n$ expansions for the relativistic and quantum electrodynamic (QED) corrections. The extrapolations based on direct calculations for the singlet and triplet $P$-states up to principal quantum number $n = 35$ provide ionization energies of the $1snp\;^1P_1$ and $^3P_c$ (centroid) states up to $n=102$ with accuracies better than $\pm$1 kHz.  The calculated ionization energies are combined with 28 measured transition frequencies to obtain values for the ionization energy of the
$1s2s\;^3S_1$ state. The final result of 1152\,842\,742.705(16) MHz differs from theory by $0.474\pm 0.052$ MHz, and provides a strong confirmation of the 9$\sigma$ disagreement between theory and experiment obtained previously by quantum defect extrapolation of experimental data to the series limit. An analysis of the quantum defect method is presented, and second-order
mass polarization (recoil) terms are identified that vary as $1/n^2$ in lowest order.  The nonrelativistic part provides a theoretical justification for the effective reduced-mass Rydberg $R_M^{(+)}$ based on the phenomenological model of a Rydberg electron scattering from a He$^+$ core.  The Ritz expansion for the nonrelativistic energy is verified to an unprecedented 20-figure accuracy. 

\end{abstract}

\date{\today}

\maketitle
\section{Introduction} 
Recent high precision measurements of transition frequencies to the high-lying singlet and triplet $P$-states of helium as high as principal quantum number $n = 102$ \cite{Clausen2021,Clausen2025} have confirmed a $9\sigma$ disagreement between theory \cite{Pachucki2017,Patkos2021} and experiment for the ionization energy of the metastable $1s2s\;^3S_1$ state.  The origin of the discrepancy remains unexplained, even though helium is a fundamental three-body problem for which high precision calculations are available, including electron correlation, relativistic and quantum electrodynamic (QED) corrections for the ground and low-lying excited states $1s2s\;^3S_1$ state \cite{Yan1995,Pachucki2012,Pachucki2010,Zheng2017,Kato2018}.  These measurements have stimulated a wave of theoretical work to develop computational methods capable of matching spectroscopic accuracy at the 1 kHz level \cite{Bondy2025,Drake2026,Chi2025,Fang2026} for the high-lying Rydberg states.  Complete calculations in Hylleraas coordinates are now available up to up to $n = 35$ \cite{Drake2026}. 

The original measurements and analysis of Clausen et al.\ \cite{Clausen2021,Clausen2025} were based on quantum defect extrapolations of the $1s2s\;^1S_0 - 1snp\;^1P_1$ and $1s2s\;^3S_1 - 1snp\;^3P_J$ transition frequencies to the series limit, using a combination of theoretical data for low-$n$ and measured transition frequencies for high-$n$ (along with the accurately known $1s2s\;^1S_0-1s2s\;^3S_1$ intercombination transition frequency \cite{Rengelink2018,Steinebach2026}), to determine the absolute ionization energy of the lower $1s2s\;^3S_1$ state. Although the quantum defect method is well-established and widely used, it is still a semi-empirical theory not fully derivable from fundamental atomic physics.  Especially the Ritz expansion, which includes only even powers of $1/(n-\delta(n))$ (where $\delta(n)$ is the quantum defect),  is questionable since relativistic and QED corrections do not satisfy the conditions for its validity \cite{DrakeAdv}.  

Our previous work circumvented the need for quantum defect extrapolations by providing direct high-precision calculations of the ionization energies of the $1snp\;^1P$ and $^3P$ states, including relativistic and QED corrections, for $n$ up to 35.  This considerably extended the range of $n$ beyond the previous limit of 10 \cite{DrakeYan92} for which high precision calculations are are available.  The extended range is made tractable in Hylleraas coordinates for the wave functions by the use of triple basis sets in which each combination of powers in the basis set members of the form
\begin{eqnarray}
\label{basis}
\phi_{i,j,k}^{(p)}({\bf r}_1,{\bf r}_2) &=& r_1^ir_2^jr_{12}^k\exp(-\alpha_pr_1-\beta_pr_2)\cos(\theta_2)\nonumber\\
                                        &&\mbox{} \pm \mbox{exchange}
\end{eqnarray}
 is included three times with independently optimized nonlinear parameters $\alpha_p$ and $\beta_p$ ($p=1,2,3$) \cite{Drake87,DrakeMakowski88}.  Here ${\bf r}_1$ and ${\bf r}_2$ are the coordinates of the two electrons relative to the nucleus, and $r_{12} = |{\bf r}_1 - {\bf r}_2|$ is the inter-electron separation. The claimed final accuracy of $\pm$1 kHz or better for $n\ge24$ is achievable because the QED contributions, (which may indeed be responsible for the discrepancy for the low-lying states), decrease in proportion to $1/n^3$, and so are suppressed by a factor of approximately 2000 at $n = 24$. The theoretical ionization energies therefore provide reliable fixed reference points for transitions to lower-lying states. 
 
 The above calculations provide reliable reference points for the 11 measured transitions for the $P$-states up to $n = 35$, but they do not provide comparison data for the remaining three triplet-$P$ measurements ($n=40,\,50\,\,55$) and 14 singlet-$P$ measurements up to $n=102$ \cite{Clausen2021,Clausen2025}. The aim of the present work is to cover this higher range up to $n=102$ by a combination of quantum defect extrapolations for the nonrelativistic energy and $1/n$ expansions for the remaining relativistic and QED corrections. This approach only makes use of QDT under conditions where it is rigorously justified, as further discussed in this article, and avoids possible complications due to relativistic and QED corrections.  The $1/n$ expansions for the latter corrections yield useful insights into the choice of parameters for a quantum defect to atomic energy levels, and especially for the choice of an effective Rydberg constant.
 
The balance of the paper is organized as follows.  Section II provides a brief summary of the overall plan of the calculation, and the computational methods used to generate the nonrelativistic wave functions in Hylleraas coordinates.  Especially important are the truncated triple basis sets needed to maintain high accuracy for the high-lying Rydberg states up to $n=35$. Section III summarizes the operators needed for the relativistic and QED contributions to the energy of order $\alpha^2$ and $\alpha^3$ Ry respectively, including relativistic recoil terms.  Section IV is the main part of the paper where quantum defect expansions are obtained for the nonrelativistic energies and $1/n$ expansions for all the other matrix elements needed to calculate the relativistic and QED contributions. This section identifies second-order mass polarization (recoil) terms that have a leading $1/n^2$ dependence with state-independent coefficients.  They provide a formal justification for the phenomenological model of an electron scattering from a He$^+$ core in calculating the reduced electron mass.  Section V presents the results for the calculated ionization energies of the Rydberg $P$-states, and their combination with the measured transition frequencies to determine the ionization energies of the metastable $2\;^1S_0$ and $2\;^3S_1$ states.  Section VI presents the conclusions and a discussion of the results.       
 
\section{Calculations}
The overall plan of the calculation is: (i) generate high-precision nonrelativistic wave functions and energies, including the effects of mass polarization, for
the helium $P$-states up to $n=35$
(ii) obtain quantum defect expansions for the nonrelativistic energies, (iii) use the wave functions to calculate 
the matrix elements needed for the relativistic and QED corrections, (iv) obtain $1/n$ expansions for the relativistic and QED corrections, and (v) use the 
combination of quantum defect expansions and $1/n$ expansions to calculate theoretical total ionization energies up to $n=102$.  Each of these steps is 
described in the following sections.

\subsection{Nonrelativistic Wave Functions and Energies}
The first computational step is to find high precision solutions to the nonrelativistic Schr\"odinger equation $H\Psi = E\Psi$ where the Hamiltonian in
center-of-mass (CM) coordinates is given by (in atomic units with $4\pi\epsilon_0 = 1$)
\begin{equation}
\label{eq:H}
H = \frac{p_1^2 + p_2^2}{2\mu} +\frac{{\bf p}_1\bdot {\bf p}_2}{M}
-\frac{Ze^2}{r_1} - \frac{Ze^2}{r_2} + \frac{e^2}{r_{12}}
\end{equation}
where the ${\bf p}_1\bdot{\bf p}_2/M$ term is the mass polarization operator resulting from the motion of the nucleus in the
CM frame, and $\mu = m_e/(M+m_e)$ is the reduced electron mass. For computational purposes, it is advantageous to transform to reduced mass atomic units where the unit of distance is the reduced mass Bohr radius $a_\mu = (m_e/m_\mu)a_0$, and the unit of energy is $e^2/a_\mu = (m_\mu/m_e)$ $E_h$.  The equation $E_{h,\mu} = (m_\mu/m_e)E_h$ defines the reduced mass hartree, and $R_M= \frac12 E_{h,\mu}$ defines the reduced mass Rydberg.  In these units, the Hamiltonian assumes the simplified form
\begin{equation}
\label{eq:Hred}
H = -\frac{1}{2}\left(\nabla_1^2 + \nabla_2^2\right) - \frac{\mu}{M}\nabla_1\bdot \nabla_2
-\frac{Z}{r_1} - \frac{Z}{r_2} + \frac{1}{r_{12}}
\end{equation}
making it clear that the solutions to the Schr\"odinger equation depend only on the ratio $\mu/M$ multiplying the mass polarization term $-\nabla_1\bdot \nabla_2$.

As discussed in detail previously \cite{Bondy2025}, the principal computational 
step is to solve the generalized matrix eigenvalue problem
\begin{equation}
{\bf\rm H}\mbox{\boldmath{$\Psi$}} =  E{\bf\rm O}\mbox{\boldmath{$\Psi$}}
\end{equation}   
where {\bf H} and {\bf O} are the Hamiltonian and overlap matrices respectively in the $\phi_{ijk}^{(p)}$ basis set, and \mbox{\boldmath{$\Psi$}} is
the column vector of wave function coefficients $c_{ijk}^{(p)}$.  Rather than by complete diagonalization, the eigenvalue $E_n$ closest to an initial guess
is found by the power method.  The corresponding wave function then has the form
\begin{equation}
\Psi({\bf r}_1,{\bf r}_2) = \sum_{p=1}^{3}\sum_{i,j,k}^{i+j+k\le\Omega_p}c_{ijk}^{(p)}\phi_{i,j,k}^{(p)}({\bf r}_1,{\bf r}_2)
\end{equation}
with $E_n$ being a variational upper bound to the true eigenvalue. The six nonlinear parameters $\alpha_p$ and $\beta_p$ are optimized by calculating analytically the six derivatives $\partial E/\partial \alpha_p$ and $\partial E/\partial \beta_p$, $p=1-3$, and using the method of steepest descent \cite{Drake87,DrakeMakowski88}. In
constructing the basis sets with $i+j+k\le\Omega_p$, numerical instability can be avoided by truncating the sequence of powers of $r_1$ and $r_{12}$ in the asysmptotic
sector $p=1$ where $\beta_1\simeq 1/(2n)$ is small \cite{Bondy2025}. 

The wave functions and energies are first generated for the hypothetical infinite nuclear mass case $M = \infty$, and then repeated with the current CODATA
value for $m_e/M(^4{\rm He}^{++}) = 1.370\,933\,554\,733(32)\times10^{-4}$ \cite{CODATA}.  Since the Hamiltonian (\ref{eq:Hred}) depends only on the ratio $\mu/M$, the exact value used is
\begin{equation}
\mu/M = \frac{m_e/M}{1+ m_e/M} = 1.370\,745\,634\,614\times10^{-4}
\end{equation}

\begingroup
\squeezetable
\begin{table*}[tb]
\caption{Nonrelativistic energies for the $1snp\;^1P$ and $1snp\;^3P$ states of helium cases of infinite nuclear mass (second and third columns in unit of h) and
the $^4$He isotope assuming $\mu/M =1.370\,745\,634\,614\times10^{-4}$ [forth and fifth columns in units of reduced mass hartrees $E_{h,\mu} = (\mu/m_e)E_h$]. }
\label{Table:nonrelE}
\begin{tabular}{rr@{}lr@{}l r@{}lr@{}l}
\hline\hline
&&\multicolumn{3}{c}{Infinite nuclear mass case}&&\multicolumn{3}{c}{Finite nuclear mass case for $^4$He}\\
 $n$  &&\multicolumn{1}{c}{  $E(n\;^1P)$ ($E_h$)} &&  \multicolumn{1}{c}{$E(n\;3P)$ ($E_h$)} &&\multicolumn{1}{c}{  $E(n\;^1P)$ ($E_{h,\mu}$)} &&  \multicolumn{1}{c}{$E(n\;3P)$ ($E_{h,\mu}$)}  \\
\hline
  2& $-2$&.123\,843\,086\,498\,101\,359\,241(22)   & $-2$&.133\,164\,190\,779\,283\,205\,140(8)    & $-2$&.123\,836\,778\,126\,6702\,663\,69(14)   & $-2$&.133\,173\,045\,867\,336\,306\,184(6)    \\
  3& $-2$&.055\,146\,362\,091\,943\,536\,93(3)     & $-2$&.058\,081\,084\,274\,275\,331\,34(3)     & $-2$&.055\,144\,369\,165\,7409\,373\,8(8)     & $-2$&.058\,083\,603\,517\,916\,261\,20(4)     \\
  4& $-2$&.031\,069\,650\,450\,240\,714\,742(12)   & $-2$&.032\,324\,354\,296\,630\,331\,956(9)    & $-2$&.031\,068\,793\,719\,9618\,396\,65(25)   & $-2$&.032\,325\,390\,598\,350\,894\,926(11)   \\
  5& $-2$&.019\,905\,989\,900\,846\,448\,848(7)    & $-2$&.020\,551\,187\,256\,267\,788\,246(11)   & $-2$&.019\,905\,547\,557\,5813\,885\,87(16)   & $-2$&.020\,551\,710\,051\,776\,754\,579(4)    \\
  6& $-2$&.013\,833\,979\,671\,740\,102\,386(9)    & $-2$&.014\,207\,958\,773\,750\,591\,219(6)    & $-2$&.013\,833\,722\,517\,7244\,983\,22(27)   & $-2$&.014\,208\,258\,474\,569\,352\,215(21)   \\
  7& $-2$&.010\,169\,314\,529\,388\,984\,898(3)    & $-2$&.010\,404\,960\,007\,971\,431\,262(5)    & $-2$&.010\,169\,152\,141\,3690\,538\,65(4)    & $-2$&.010\,405\,147\,524\,909\,791\,537(7)    \\
  8& $-2$&.007\,789\,127\,133\,235\,895\,156(3)    & $-2$&.007\,947\,013\,771\,161\,505\,262(5)    & $-2$&.007\,789\,018\,149\,3779\,248\,16(6)    & $-2$&.007\,947\,138\,808\,412\,996\,650(14)   \\
  9& $-2$&.006\,156\,384\,652\,853\,818\,346(5)    & $-2$&.006\,267\,267\,366\,409\,032\,580(14)   & $-2$&.006\,156\,308\,015\,2488\,294\,02(8)    & $-2$&.006\,267\,354\,877\,217\,267\,49(6)     \\
 10& $-2$&.004\,987\,983\,802\,218\,239\,45818(3)  & $-2$&.005\,068\,805\,497\,707\,316\,44544(3)  & $-2$&.004\,987\,927\,884\,4113\,185\,468(1)   & $-2$&.005\,068\,869\,120\,520\,818\,9045(1)   \\
 11& $-2$&.004\,123\,191\,922\,332\,652\,52917(3)  & $-2$&.004\,183\,903\,199\,590\,642\,37633(2)  & $-2$&.004\,123\,149\,883\,7746\,324\,9720(1)  & $-2$&.004\,183\,950\,897\,840\,245\,9467(1)   \\
 12& $-2$&.003\,465\,252\,704\,885\,798\,26035(5)  & $-2$&.003\,512\,006\,535\,142\,745\,5602(2)   & $-2$&.003\,465\,220\,309\,3993\,091\,715(1)   & $-2$&.003\,512\,043\,211\,271\,315\,0455(1)   \\
 13& $-2$&.002\,953\,093\,958\,149\,784\,87063(6)  & $-2$&.002\,989\,859\,764\,908\,816\,32221(5)  & $-2$&.002\,953\,068\,469\,4612\,088\,6054(2)  & $-2$&.002\,989\,888\,570\,652\,384\,6102(1)   \\
 14& $-2$&.002\,546\,625\,370\,190\,968\,01192(1)  & $-2$&.002\,576\,056\,426\,625\,769\,0486(3)   & $-2$&.002\,546\,604\,957\,3601\,277\,470(1)   & $-2$&.002\,576\,079\,462\,858\,174\,6833(1)   \\
 15& $-2$&.002\,218\,647\,104\,088\,301\,58826(6)  & $-2$&.002\,242\,571\,222\,150\,326\,43884(5)  & $-2$&.002\,218\,630\,504\,6558\,117\,5390(1)  & $-2$&.002\,242\,589\,932\,872\,481\,8034(1)   \\
 16& $-2$&.001\,950\,177\,973\,979\,340\,789(10)   & $-2$&.001\,969\,887\,403\,296\,966\,97718(7)  & $-2$&.001\,950\,164\,294\,6972\,926\,7779(3)  & $-2$&.001\,969\,902\,807\,498\,563\,6564(1)   \\
 17& $-2$&.001\,727\,645\,999\,910\,516\,9444(8)   & $-2$&.001\,744\,075\,191\,087\,356\,0050(2)   & $-2$&.001\,727\,634\,594\,4149\,818\,42(6)    & $-2$&.001\,744\,088\,024\,464\,174\,062(5)    \\
 18& $-2$&.001\,541\,138\,760\,913\,624\,7085(1)   & $-2$&.001\,554\,976\,940\,232\,726\,65182(4)  & $-2$&.001\,541\,129\,152\,1801\,259\,446(2)   & $-2$&.001\,554\,987\,744\,693\,228\,7005(4)   \\
 19& $-2$&.001\,383\,280\,102\,341\,609\,78060(2)  & $-2$&.001\,395\,044\,604\,295\,124\,98775(1)  & $-2$&.001\,383\,271\,932\,1276\,488\,8003(1)  & $-2$&.001\,395\,053\,786\,133\,765\,122(5)    \\
 20& $-2$&.001\,248\,489\,450\,687\,613\,1089(1)   & $-2$&.001\,258\,574\,692\,820\,423\,80893(9)  & $-2$&.001\,248\,482\,445\,7419\,142\,446(1)   & $-2$&.001\,258\,582\,561\,493\,061\,8850(2)   \\
 21& $-2$&.001\,132\,481\,733\,017\,529\,51738(2)  & $-2$&.001\,141\,192\,656\,477\,678\,69139(3)  & $-2$&.001\,132\,475\,681\,9913\,937\,5668(8)  & $-2$&.001\,141\,199\,451\,033\,505\,703(2)    \\
 22& $-2$&.001\,031\,922\,552\,162\,285\,43555(1)  & $-2$&.001\,039\,497\,912\,816\,701\,20993(1)  & $-2$&.001\,031\,917\,289\,5185\,297\,0734(3)  & $-2$&.001\,039\,503\,820\,281\,064\,6260(2)   \\
 23& $-2$&.000\,944\,185\,877\,955\,806\,44396(1)  & $-2$&.000\,950\,814\,760\,962\,146\,31469(6)  & $-2$&.000\,944\,181\,272\,5530\,152\,605(11)  & $-2$&.000\,950\,819\,929\,354\,245\,5031(2)   \\
 24& $-2$&.000\,867\,180\,846\,170\,111\,28223(6)  & $-2$&.000\,873\,014\,566\,616\,659\,39240(9)  & $-2$&.000\,867\,176\,793\,0321\,248\,2530(4)  & $-2$&.000\,873\,019\,114\,319\,712\,0010(3)   \\
 25& $-2$&.000\,799\,226\,024\,103\,063\,04556(1)  & $-2$&.000\,804\,386\,829\,929\,070\,60845(2)  & $-2$&.000\,799\,222\,438\,3972\,552\,64(11)   & $-2$&.000\,804\,390\,852\,535\,846\,8347(1)   \\
 26& $-2$&.000\,738\,956\,837\,741\,719\,21765(4)  & $-2$&.000\,743\,544\,360\,330\,295\,8128(2)   & $-2$&.000\,738\,953\,650\,3138\,422\,89(6)    & $-2$&.000\,743\,547\,935\,712\,491\,34(11)    \\
 27& $-2$&.000\,685\,256\,528\,882\,402\,12145(2)  & $-2$&.000\,689\,352\,623\,570\,976\,55876(3)  & $-2$&.000\,685\,253\,682\,9046\,670\,99(14)   & $-2$&.000\,689\,355\,815\,680\,215\,8588(5)   \\
 28& $-2$&.000\,637\,204\,047\,170\,837\,14304(3)  & $-2$&.000\,640\,876\,467\,041\,024\,09302(2)  & $-2$&.000\,637\,201\,495\,5970\,293\,5(17)    & $-2$&.000\,640\,879\,328\,791\,590\,5337(6)   \\
 29& $-2$&.000\,594\,034\,290\,981\,601\,7891(1)   & $-2$&.000\,597\,339\,504\,545\,631\,8545(1)   & $-2$&.000\,594\,031\,994\,5946\,068\,210(1)   & $-2$&.000\,597\,342\,080\,024\,142\,5139(4)   \\
 30& $-2$&.000\,555\,107\,462\,372\,974\,2591(24)  & $-2$&.000\,558\,092\,835\,757\,975\,2832(5)   & $-2$&.000\,555\,105\,388\,2722\,436\,086(3)   & $-2$&.000\,558\,095\,161\,933\,396\,4815(7)   \\
 31& $-2$&.000\,519\,885\,224\,314\,856\,9889(1)   & $-2$&.000\,522\,590\,726\,604\,946\,1686(3)   & $-2$&.000\,519\,883\,344\,7339\,817\,407(3)   & $-2$&.000\,522\,592\,834\,664\,469\,519(2)    \\
 32& $-2$&.000\,487\,911\,987\,756\,799\,3241(1)   & $-2$&.000\,490\,371\,534\,947\,756\,0339(3)   & $-2$&.000\,487\,910\,279\,1315\,492\,9(25)    & $-2$&.000\,490\,373\,451\,349\,027\,22(23)    \\
 33& $-2$&.000\,458\,800\,104\,867\,467\,8590(1)   & $-2$&.000\,461\,042\,627\,369\,818\,5858(5)   & $-2$&.000\,458\,798\,547\,0952\,679\,93(13)   & $-2$&.000\,461\,044\,374\,676\,264\,00(5)     \\
 34& $-2$&.000\,432\,218\,063\,626\,956\,6102(1)   & $-2$&.000\,434\,268\,360\,440\,676\,5523(1)   & $-2$&.000\,432\,216\,639\,4850(14)            & $-2$&.000\,434\,269\,957\,990\,4(11)          \\
 35& $-2$&.000\,407\,881\,008\,092\,820\,967(1)  & $-2$&.000\,409\,760\,435\,014\,704\,07(22)      & $-2$&.000\,407\,879\,702\,7180\,4(16)         & $-2$&.000\,409\,761\,899\,456\,37(4)          \\
\hline
\hline
\end{tabular}
\end{table*} 
\endgroup 
  
Results for the nonrelativistic energies for both the infinite and finite mass cases are summarized for completeness  in Table \ref{Table:nonrelE}.  Energies for the infinite mass case were tabulated previously \cite{Drake2026} and compared with the correlated B-spline results of Chi et al.\ \cite{Chi2025} and Fang et al.\ \cite{Fang2026}.  The present values are of improved accuracy for the two highest values $n = 34$ and 35.  A uniform accuracy of approximately 21 significant figures in the energy
is obtained by progressively increasing $\Omega_1$ from 18 for $n = 2$ (2997 terms) to 37 for $n = 35$ (8588 terms).  Results for the finite mass case have not been tabulated previously.  They are of comparable accuracy, except possibly for $n=35$, where the accuracy begins to deteriorate.  Standard double precision arithmetic (approximately 32 decimal digits) is adequate up to $n \simeq 16$.  After that, Bailey's
double-quadruple (dq) precision arithmetic package \cite{Bailey} (approximately 70 decimal digits) is used.   A complete calculation for the largest 8588-term basis set in dq precision takes approximately 20 hours on a workstation with a single cpu, including the calculation of derivatives
$\partial E/\partial \alpha_p$ etc.  

\section{Relativistic and QED Corrections}
\label{relativistic}
the next step is to calculate the relativistic and QED corrections as expectation values using the standard methods of nrQED. 
To order $\alpha^2$ Ry ($\alpha^4mc^2$), the leading terms are the well-known Breit interaction terms (superscripts denote the power of $\alpha$)
(in atomic units) \cite{DrakeYan92,Bethe}
\begin{equation}
E_{\rm rel}^{(2)} = \sum_{i=1}^5 \langle B_i\rangle + \frac{m_e}{M}(\tilde\Delta_2 + \tilde\Delta_{3Z})
\end{equation}
where $B_1 = -(p_1^4+p_2^4)/(8m^3c^2)$ arises from the relativistic variation of mass with velocity, $B_2$ is the orbit-orbit interaction defined by
\begin{equation}
B_2 = -\frac{e^2}{2(mc)^2}\left[\frac{{\bf p}_1\bdot {\bf p}_2}{r_{12}} + \frac{{\bf r}_{12}\bdot({\bf r}_{12}\bdot{\bf p}_1){\bf p_2}}{r_{12}^3}\right]
\end{equation}
$B_3 = B_{\rm so} + B_{\rm soo}$ contains the spin-orbit and spin-other-orbit interactions defined by
\begin{eqnarray}
B_{\rm so} &=& \frac{Z\mu_{\rm B}e}{mc}\left[\frac{{\bf r}_1\times{\bf p}_1\bdot{\bf s}_1}{r_1^3} + \frac{{\bf r}_2\times{\bf p}_2\bdot{\bf s}_2}{r_2^3}\right]\\
B_{\rm soo} &=& -\frac{\mu_{\rm B}e}{2mcr_{12}^3}{\bf r}_{12}\times\left[3{\bf p}_-\bdot{\bf s}_+ -
{\bf p}_+\bdot{\bf s}_-\right]
\end{eqnarray}
where ${\bf p}_\pm = {\bf p}_1\pm{\bf p_2}$, ${\bf s}_\pm = {\bf s}_1\pm{\bf s_2}$, $\mu_{\rm B} = e\hbar/(2mc)$ is the Bohr magneton.
The anomalous magnetic moment term defined by
\begin{equation}
B_{\rm anom} = a_e(2B_{\rm so} + \textstyle\frac43B_{\rm soo})
\end{equation}
is calculated and added in separately, where $a_{\rm e} \simeq\alpha/(2\pi) -0.328\,429\alpha^2$ is the electron anomalous magnetic moment.
$B_5$ is the spin-spin term $B_{\rm ss}$ defined by
\begin{equation}
B_{\rm ss} = 4\mu_e^2\left[\frac{8\pi}{3}\delta(r_{12}){\bf s}_1\bdot{\bf s}_2+
\frac{{\bf s}_1\bdot{\bf s}_2}{r_{12}^3} - \frac{3({\bf s}_1\bdot{\bf r}_{12}){\bf s}_2\bdot{\bf r}_{12}}{r_{12}^5}\right],
\end{equation}
where $\mu_e = \mu_{\rm B}(1 + a_e)$, and the relativistic recoil terms are
\begin{eqnarray}
\tilde\Delta_2&=&-\frac{Z e^2}{2(mc)^2} \sum_{j=1} ^2 \left[\frac{1}{r_j}{
\bf{p_+}} \bdot {\bf p}_j +\frac{1}{r_j^3}{\bf{r}}_j \bdot ({\bf{r}}
_j \bdot {\bf{p_+}}){\bf p}_j \right],\\
\label{eq:009}
\tilde\Delta_{3Z}&=&\frac{2Z\mu_{\rm B}e}{mc}\sum_{i=1} ^2 \frac{1}{r_i^3}{\bf{r}}
_i \times {\bf{p_+}} \bdot {\bf{s}}_i.
\label{eq:010}
\end{eqnarray}
  Finally
 $B_4$ contains the $\delta$-function terms $B_4 = \alpha^2\pi[\frac12\delta(r_1)
+ \frac12\delta(r_2) - \delta(r_{12})]$.  To each of these, there are finite nuclear mass corrections of order $\alpha^2\mu/M$ arising from (i) the mass scaling of each term, and (ii) cross terms between the $B_i$ terms in $E_{\rm rel}^{(2)}$ and the mass polarization operator in Eq.\ (\ref{eq:H}).
The relativistic recoil terms $\tilde\Delta_2$ and $\tilde\Delta_{3Z}$ arise from the transformation of the Breit interaction to CM plus relative coordinates \cite{DrakeYan92,Stone1963}.
In the material to follow, the mass scaling and recoil terms taken together are denoted by $E_{\rm RR,M}^{(2)}$, and the cross terms (ii) by $E_{\rm RR,X}^{(2)}$.

The leading QED corrections of order $\alpha^3$ Ry ($\alpha^5mc^2)$ due to electron self-energy and vacuum polarization can be divided into an electron-nucleus part ($E_{\rm L,1}^{(3)}$) \cite{Kabir} and an electron-electron part ($E_{\rm L,2}^{(3)}$) \cite{Araki,Sucher} defined by
\begin{eqnarray}
 E_{\rm L,1}^{(3)} =&&\!\!\!\!\! \textstyle\case43Z\alpha^3[\case{19}{30} -\ln(Z\alpha)^2 - \ln k_0]\langle\delta(r_1) + \delta(r_2)\rangle\;\;\\
 E_{\rm L,2}^{(3)} =&&\!\!\!\!\! \textstyle\alpha^3[(\case{89}{15} + \case{14}{3}\ln\alpha - \case{20}{3}{\bf s}_1\bdot{\bf s}_2)\langle\delta(r_{12})\rangle-\case{14}{3}Q]
\end{eqnarray}
where $k_0$ is Bethe's mean excitation energy \cite{Bethe} in units of $Z^2$ Ry.
The $\ln Z^2$ scaling of the Bethe logarithm is included in the $\ln(Z\alpha)^2$ term so that the value is then close to  the hydrogenic value 2.984\,128\,556 ($\pm$0.5\%) for all states of all light atoms and ions studied \cite{YanDrake2003,PachuckiKomasa2004,Korobov2019,Lesiuk2024}.  Its value for the Rydberg $^1P$ and $^3P$ states of helium can be accurately estimated from the $1/n$ expansions of Drake \cite{Drake2001} or Korobov \cite{Korobov2019}.

 For $\Delta E_{\rm L,2}^{(3)}$, the $Q$ term
is defined by the improper integral
\begin{equation}
\label{Q0}
Q = \frac{1}{4\pi}\lim_{\epsilon\rightarrow 0}\langle
r_{12}^{-3}(\epsilon)+ 4 \pi (\gamma +\ln \epsilon) \delta({\bf r}_{12})\rangle\,
\end{equation}
where $\epsilon$ is the radius of an infinitesimal sphere that is excluded from the range of integration, and $\gamma$ is Euler's constant.

Contributions of order $\alpha^4$ Ry have been
calculated in their entirety for the $2\,^1P_1$ state by Pachucki et al.\
\cite{Pachucki2017}.  Our strategy is to calculate the dominant parts that can be easily evaluated, subtract these from the results in Ref.\ \cite{Pachucki2017}, and estimate the remainder from its approximately $1/n^3$ scaling with $n$, as discussed in detail previously (see Table III of Ref.\ \cite{Drake2026}).  The estimated remainders $E^{(4)}_{\rm rmdr}$ are included in the final results with the entire amount taken as the uncertainty. As a check, the scaling argument applied to the singlet-triplet mixing term $E^{(4)}_{{\rm st},n}$ reduces its value by a factor of 2460 for $n=27$ to 1.93 kHz, which is close to the correct value 1.60 kHz (see Table IV of Ref.\ \cite{Drake2026}).  

The last correction to be included is that due to finite nuclear size.  To lowest order, the energy shift to sufficient accuracy is $E_{\rm nuc} = \frac23\pi Z(R/a_0)^2\langle\delta({\bf r}_1) + \delta({\bf r}_2)\rangle$ where $a_0$ is the Bohr radius and $R=1.6786(12)$ fm \cite{Pachucki2024} is the radius of the nuclear charge distribution for $^4$He.
\begingroup
\squeezetable
\begin{table*}[tb]
\caption{Matrix elements scaled by $n^3$ needed to calculate the relativistic and QED contributions to the energy for the $1snp\;^1P$ states of helium
for the case of infinite nuclear mass. For $\langle p^4\rangle$, $\pi\langle\delta(\br_1)\rangle$ and ($\langle Q_1\rangle$, the hydrogenic $1s$ electron
contribution is subtracted before scaling so that the remainder asymptotically tends to a constant. Unit are atomic units.}
\label{Table:singlet0}
\begin{tabular}{rr@{}lr@{}lr@{}lr@{}lr@{}lr@{}lr@{}l}
\hline\hline
 $n$  &&$n^3(\langle p_1^4\rangle-40)$ &&$n^3\langle B_2\rangle$ &&$n^3(\pi\langle\delta(\br_1)\rangle - 4)$&&$n^3\pi\langle\delta(\br_{12}\rangle$ &&      $n^3\langle\tilde\Delta_2\rangle$ &&$n^3\langle Q\rangle$&&$n^3(\langle Q_1\rangle-\frac{16}{\pi}\ln2)$\\
\hline
  2 & -0&.234\,010\,8868(2)& -0&.162\,643\,7927& 0&.028\,986\,306(3) &0&.018\,476\,8092   & -2&.292\,032\,3673(4)&0&.026\,995\,9790  & -7&.060\,581\,7829(1) \\
  3 & -0&.307\,066\,2564   & -0&.180\,828\,9600& 0&.032\,889\,7834(5)&0&.021\,376\,6192   & -2&.746\,190\,6114   &0&.026\,922\,7377   & -7&.060\,544\,5847(1)\\
  4 & -0&.340\,232\,6974(1)& -0&.186\,769\,8115& 0&.033\,427\,6335(6)&0&.022\,411\,1758   & -2&.951\,264\,1224(1)&0&.026\,944\,9469(1)& -7&.060\,432\,8618(1)\\
  5 & -0&.359\,713\,8701(3)& -0&.189\,510\,5317& 0&.033\,552\,330(3) &0&.022\,900\,9286(1)& -3&.069\,457\,2602(8)&0&.026\,969\,7913(2)& -7&.060\,387\,7852(1)\\
  6 & -0&.372\,624\,5880(4)& -0&.191\,021\,7175& 0&.033\,592\,050(1) &0&.023\,172\,9001(1)& -3&.146\,548\,6985(7)&0&.026\,990\,2397(1)& -7&.060\,367\,3649   \\
  7 & -0&.381\,832\,1397(1)& -0&.191\,952\,7024& 0&.033\,608\,6272(3)&0&.023\,340\,4136   & -3&.200\,852\,5033(2)&0&.027\,006\,5431   & -7&.060\,356\,9251   \\
  8 & -0&.388\,737\,0736   & -0&.192\,571\,6542& 0&.033\,617\,548(1) &0&.023\,451\,3883   & -3&.241\,185\,6587   &0&.027\,019\,6340   & -7&.060\,351\,0623   \\
  9 & -0&.394\,109\,802(2) & -0&.193\,006\,7432& 0&.033\,623\,558(5) &0&.023\,528\,9962   & -3&.272\,329\,740(4) &0&.027\,030\,3022   & -7&.060\,347\,5238   \\
 10 & -0&.398\,410\,4764   & -0&.193\,325\,8997& 0&.033\,628\,2412   &0&.023\,585\,5874   & -3&.297\,105\,5823(1)&0&.027\,039\,1319   & -7&.060\,345\,2640   \\
 11 & -0&.401\,931\,3508   & -0&.193\,568\,0265& 0&.033\,632\,190(2) &0&.023\,628\,2498(1)& -3&.317\,285\,8619(1)&0&.027\,046\,5456(1)& -7&.060\,343\,7539   \\
 12 & -0&.404\,867\,1375(1)& -0&.193\,756\,7897& 0&.033\,635\,6848(3)&0&.023\,661\,2959   & -3&.334\,040\,9246(2)&0&.027\,052\,8508   & -7&.060\,342\,7067   \\
 13 & -0&.407\,352\,6293(1)& -0&.193\,907\,3019& 0&.033\,638\,8518   &0&.023\,687\,4762   & -3&.348\,174\,7031(1)&0&.027\,058\,2743   & -7&.060\,341\,9579   \\
 14 & -0&.409\,484\,1060(1)& -0&.194\,029\,6063& 0&.033\,641\,7369   &0&.023\,708\,6148   & -3&.360\,257\,7058(1)&0&.027\,062\,9865   & -7&.060\,341\,4083   \\
 15 & -0&.411\,332\,219(3) & -0&.194\,130\,6036& 0&.033\,644\,388(3) &0&.023\,725\,9621   & -3&.370\,706\,089(5) &0&.027\,067\,1170   & -7&.060\,340\,9958   \\
 16 & -0&.412\,949\,981(3) & -0&.194\,215\,1719& 0&.033\,646\,840(4) &0&.023\,740\,3992   & -3&.379\,830\,539(7) &0&.027\,070\,7663   & -7&.060\,340\,6802   \\
 17 & -0&.414\,377\,9487(1)& -0&.194\,286\,8444& 0&.033\,649\,1125(6)&0&.023\,752\,5622   & -3&.387\,867\,6851(5)&0&.027\,074\,0132   & -7&.060\,340\,4346   \\
 18 & -0&.415\,647\,6774(1)& -0&.194\,348\,2356& 0&.033\,651\,222(1) &0&.023\,762\,9206   & -3&.395\,000\,9338(1)&0&.027\,076\,9202   & -7&.060\,340\,2407   \\
 19 & -0&.416\,784\,0969(1)& -0&.194\,401\,3162& 0&.033\,653\,1837(6)&0&.023\,771\,8267   & -3&.401\,374\,6432(5)&0&.027\,079\,5378   & -7&.060\,340\,0855   \\
 20 & -0&.417\,807\,1550   & -0&.194\,447\,5966& 0&.033\,655\,0069   &0&.023\,779\,5498   & -3&.407\,103\,9766   &0&.027\,081\,9069   & -7&.060\,339\,9598   \\
 21 & -0&.418\,733\,016(2) & -0&.194\,488\,2514& 0&.033\,656\,714(2) &0&.023\,786\,2985   & -3&.412\,281\,954(4) &0&.027\,084\,0610   & -7&.060\,339\,8569   \\
 22 & -0&.419\,574\,9044(6)& -0&.194\,524\,2065& 0&.033\,658\,3100(2)&0&.023\,792\,2366   & -3&.416\,984\,502(1) &0&.027\,086\,0282   & -7&.060\,339\,7719   \\
 23 & -0&.420\,343\,747(2) & -0&.194\,556\,2001& 0&.033\,659\,803(3) &0&.023\,797\,4944   & -3&.421\,274\,215(3) &0&.027\,087\,8315   & -7&.060\,339\,7011   \\
 24 & -0&.421\,048\,6608(8)& -0&.194\,584\,8274& 0&.033\,661\,208(1) &0&.023\,802\,1766   & -3&.425\,203\,177(6) &0&.027\,089\,4907   & -7&.060\,339\,6416   \\
 25 & -0&.421\,697\,2968(6)& -0&.194\,610\,5731& 0&.033\,662\,5251(3)&0&.023\,806\,3679   & -3&.428\,815\,037(1) &0&.027\,091\,0222   & -7&.060\,339\,5913   \\
 26 & -0&.422\,296\,1389   & -0&.194\,633\,8351& 0&.033\,663\,7648(1)&0&.023\,810\,1378   & -3&.432\,146\,72(1)  &0&.027\,092\,4402   & -7&.060\,339\,5484   \\
 27 & -0&.422\,850\,7103(8)& -0&.194\,654\,9434& 0&.033\,664\,9355(3)&0&.023\,813\,5438   & -3&.435\,229\,4(2)   &0&.027\,093\,7568   & -7&.060\,339\,5116   \\
 28 & -0&.423\,365\,7434(8)& -0&.194\,674\,1731& 0&.033\,666\,0388(1)&0&.023\,816\,6336   & -3&.438\,090\,488(1) &0&.027\,094\,9825   & -7&.060\,339\,4799   \\
 29 & -0&.423\,845\,3199(7)& -0&.194\,691\,7555& 0&.033\,667\,076(1) &0&.023\,819\,4471   & -3&.440\,752\,591(2) &0&.027\,096\,1264   & -7&.060\,339\,4524   \\
 30 & -0&.424\,292\,9842(2)& -0&.194\,707\,8864& 0&.033\,668\,069(3) &0&.023\,822\,0180   & -3&.443\,235\,9146(4)&0&.027\,097\,1965   & -7&.060\,339\,4284   \\
 31 & -0&.424\,711\,8163(1)& -0&.194\,722\,7325& 0&.033\,669\,0043(4)&0&.023\,824\,3750   & -3&.445\,557\,8779   &0&.027\,098\,1995   & -7&.060\,339\,4075   \\
 32 & -0&.425\,104\,5152   & -0&.194\,736\,4362& 0&.033\,669\,8917   &0&.023\,826\,5425   & -3&.447\,734\,(3)    &0&.027\,099\,1416   & -7&.060\,339\,3891   \\
 33 & -0&.425\,473\,412(3) & -0&.194\,749\,1204& 0&.033\,670\,737(5) &0&.023\,828\,5413   & -3&.449\,776\,791(3) &0&.027\,100\,0282   & -7&.060\,339\,3728   \\
 34 & -0&.425\,82(1)       & -0&.194\,760\,8912& 0&.033\,671\,538(7) &0&.023\,830\,3897(1)& -3&.451\,698\,898(5) &0&.027\,100\,8640(1)& -7&.060\,339\,3585   \\
 35 & -0&.425\,87(2)       & -0&.194\,771\,8407& 0&.033\,672\,3024(4)&0&.023\,832\,1031   & -3&.453\,510\,4734(4)&0&.027\,101\,6533   & -7&.060\,339\,3457   \\
\hline
\hline
\end{tabular}
\end{table*}
\endgroup

\begin{table*}[tb]
\caption{Matrix elements scaled by $n^3$ needed to calculate the relativistic and QED contributions to the energy for the $1snp\;^3P$ states of helium
for the case of infinite nuclear mass. For $\langle p^4\rangle$, $\pi\langle\delta(\br_1)\rangle$ and ($\langle Q_1\rangle$, the hydrogenic $1s$ electron
contribution is subtracted before scaling so that the remainder asymptotically tends to a constant. Unit are atomic units.}
\label{Table:triplet0}
\begin{tabular}{rr@{}lr@{}lr@{}lr@{}lr@{}lr@{}lr@{}l}
\hline\hline
 $n$  &&$n^3(\langle p_1^4\rangle-40)$ &&$n^3\langle B_2\rangle$ &&$n^3(\pi\langle\delta(\br_1)\rangle - 4)$&&$n^3\pi\langle\delta(\br_{12}\rangle$ &&      $n^3\langle\tilde\Delta_2\rangle$ &&$n^3\langle Q\rangle$&&$n^3(\langle Q_1\rangle-\frac{16}{\pi}\ln2)$\\
\hline
  2    &0&.703\,250\,2484(1)   &0&.280\,647\,0947    & -0&.361\,382\,163(1)    &0&.0      &1&.930\,542\,2772(2)   &0&.030\,511\,3433    & -7&.013\,457\,2207(4)\\
  3    &0&.591\,888\,477(1)    &0&.279\,307\,3369    & -0&.349\,279\,507(6)    &0&.0      &1&.495\,838\,656(3)    &0&.028\,009\,6660    & -7&.046\,880\,0203(6)\\
  4    &0&.538\,890\,1653(2)   &0&.275\,797\,5672    & -0&.342\,076\,8289(3)   &0&.0      &1&.276\,636\,5726(3)   &0&.027\,014\,5648    & -7&.054\,772\,4599   \\
  5    &0&.508\,903\,2128      &0&.273\,309\,8578    & -0&.337\,821\,5692(1)   &0&.0      &1&.149\,197\,8311      &0&.026\,499\,8234    & -7&.057\,523\,0998   \\
  6    &0&.489\,754\,4524(1)   &0&.271\,563\,2872    & -0&.335\,062\,5878      &0&.0      &1&.066\,456\,7472(2)   &0&.026\,190\,1313    & -7&.058\,722\,3879   \\
  7    &0&.476\,503\,8181      &0&.270\,288\,9565    & -0&.333\,140\,708(2)    &0&.0      &1&.008\,543\,2941(1)   &0&.025\,985\,0066    & -7&.059\,326\,7093   \\
  8    &0&.466\,802\,5578(4)   &0&.269\,323\,6404    & -0&.331\,729\,182(3)    &0&.0      &0&.965\,784\,3438(7)   &0&.025\,839\,8386    & -7&.059\,663\,7143   \\
  9    &0&.459\,398\,1779(7)   &0&.268\,569\,0785    & -0&.330\,650\,1483(4)   &0&.0      &0&.932\,936\,646(1)    &0&.025\,732\,0191    & -7&.059\,866\,2950   \\
 10    &0&.453\,563\,7461(2)   &0&.267\,963\,8830    & -0&.329\,799\,2723(2)   &0&.0      &0&.906\,919\,5331(3)   &0&.025\,648\,9448    & -7&.059\,995\,3229   \\
 11    &0&.448\,849\,0645(4)   &0&.267\,468\,0942    & -0&.329\,111\,514(6)    &0&.0      &0&.885\,806\,7226(8)   &0&.025\,583\,0656    & -7&.060\,081\,3702   \\
 12    &0&.444\,960\,6783      &0&.267\,054\,7079    & -0&.328\,544\,2799(6)   &0&.0      &0&.868\,332\,8409(1)   &0&.025\,529\,5968    & -7&.060\,140\,9430   \\
 13    &0&.441\,699\,2544(5)   &0&.266\,704\,8692    & -0&.328\,068\,553(1)    &0&.0      &0&.853\,632\,799(1)    &0&.025\,485\,3657    & -7&.060\,183\,4888   \\
 14    &0&.438\,924\,721(1)    &0&.266\,405\,0361    & -0&.327\,663\,934(2)    &0&.0      &0&.841\,095\,331(2)    &0&.025\,448\,1891    & -7&.060\,214\,6814   \\
 15    &0&.436\,535\,770(3)    &0&.266\,145\,2417    & -0&.327\,315\,636(4)    &0&.0      &0&.830\,276\,299(7)    &0&.025\,416\,5172    & -7&.060\,238\,0688   \\
 16    &0&.434\,457\,3405(5)   &0&.265\,917\,9931    & -0&.327\,012\,6869(5)   &0&.0      &0&.820\,845\,296(1)    &0&.025\,389\,2206    & -7&.060\,255\,9475   \\
 17    &0&.432\,632\,6493      &0&.265\,717\,5514    & -0&.326\,746\,7934(1)   &0&.0      &0&.812\,551\,4659(1)   &0&.025\,365\,4575    & -7&.060\,269\,8484   \\
 18    &0&.431\,017\,954(1)    &0&.265\,539\,4485    & -0&.326\,511\,568(2)    &0&.0      &0&.805\,200\,945(2)    &0&.025\,344\,5877    & -7&.060\,280\,8191   \\
 19    &0&.429\,579\,008(3)    &0&.265\,380\,1545    & -0&.326\,302\,003(4)    &0&.0      &0&.798\,641\,534(6)    &0&.025\,326\,1163    & -7&.060\,289\,5927   \\
 20    &0&.428\,288\,6222(2)   &0&.265\,236\,8447    & -0&.326\,114\,1253(9)   &0&.0      &0&.792\,752\,0929(6)   &0&.025\,309\,6546    & -7&.060\,296\,6929   \\
 21    &0&.427\,124\,933(2)    &0&.265\,107\,2320    & -0&.325\,944\,736(4)    &0&.0      &0&.787\,434\,996(4)    &0&.025\,294\,8932    & -7&.060\,302\,5003   \\
 22    &0&.426\,070\,1698(2)   &0&.264\,989\,4450    & -0&.325\,791\,2411(1)   &0&.0      &0&.782\,610\,7113(3)   &0&.025\,281\,5829    & -7&.060\,307\,2963   \\
 23    &0&.425\,109\,7303(1)   &0&.264\,881\,9379    & -0&.325\,651\,507(2)    &0&.0      &0&.778\,213\,7772(1)   &0&.025\,269\,5206    & -7&.060\,311\,2918   \\
 24    &0&.424\,231\,5109(2)   &0&.264\,783\,4228    & -0&.325\,523\,7606(2)   &0&.0      &0&.774\,197\,(7)       &0&.025\,258\,5396    & -7&.060\,314\,6472   \\
 25    &0&.423\,425\,3926      &0&.264\,692\,8180    & -0&.325\,406\,5281      &0&.0      &0&.770\,493\,3639(1)   &0&.025\,248\,5011    & -7&.060\,317\,4855   \\
 26    &0&.422\,682\,8514(4)   &0&.264\,609\,2084    & -0&.325\,298\,5649(6)   &0&.0      &0&.767\,086\,8(9)      &0&.025\,239\,2894    & -7&.060\,319\,9028   \\
 27    &0&.421\,996\,6543      &0&.264\,531\,8147    & -0&.325\,198\,8122(5)   &0&.0      &0&.763\,935\,0103      &0&.025\,230\,8066    & -7&.060\,321\,9743   \\
 28    &0&.421\,360\,6256(5)   &0&.264\,459\,9692    & -0&.325\,106\,3694(9)   &0&.0      &0&.761\,012\,606(1)    &0&.025\,222\,9699    & -7&.060\,323\,7596   \\
 29    &0&.420\,769\,455(2)    &0&.264\,393\,0964    & -0&.325\,020\,4612      &0&.0      &0&.758\,294\,752(3)    &0&.025\,215\,7084    & -7&.060\,325\,3066   \\
 30    &0&.420\,218\,4(2)      &0&.264\,330\,6975    & -0&.324\,940\,4229(2)   &0&.0      &0&.755\,760\,729(4)    &0&.025\,208\,9611    & -7&.060\,326\,6537   \\
 31    &0&.419\,703\,9729(8)   &0&.264\,272\,3385    & -0&.324\,865\,671(2)    &0&.0      &0&.753\,392\,461(3)    &0&.025\,202\,6755    & -7&.060\,327\,8322   \\
 32    &0&.419\,227(4)       &0&.264\,217\,6395    & -0&.324\,795\,6951(3)   &0&.0      &0&.751\,174\,2291(1)   &0&.025\,196\,8059    & -7&.060\,328\,8676   \\
 33    &0&.418\,745(3)       &0&.264\,166\,2669    & -0&.324\,730\,0583(1)   &0&.0      &0&.749\,092\,2071(4)   &0&.025\,191\,3124    & -7&.060\,329\,7811   \\
 34    &0&.418\,36(2)        &0&.264\,117\,9264    & -0&.324\,668\,36(1)     &0&.0      &0&.747\,134\,217(6)    &0&.025\,186\,1601    & -7&.060\,330\,5900   \\
 35    &0&.419\,49(2)        &0&.264\,072\,3556    & -0&.324\,610\,255(9)    &0&.0      &0&.745\,289\,497(9)    &0&.025\,181\,3183    & -7&.060\,331\,3089   \\
\hline
\hline
\end{tabular}
\end{table*}

\begin{table*}[tb]
\caption{Mass polarization coefficients $X^{(1)}$ scaled by $n^3$ for matrix elements needed to calculate the relativistic and QED contributions to the energy for the $1snp\;^1P$ states of helium. See Eqs.\ (\ref{eq:X(1)}) and (\ref{eq:X(M)}).  Unit are atomic units.}
\label{Table:singletM}
\begin{tabular}{rr@{}lr@{}lr@{}lr@{}lr@{}lr@{}lr@{}l}
\hline\hline
 $n$  &&$n^3\langle p_1^4\rangle^{(1)}$ &&$n^3\langle B_2\rangle^{(1)}$ &&$n^3\pi\langle\delta(\br_1)\rangle^{(1)}$&&$n^3\pi\langle\delta(\br_{12}\rangle^{(1)}$ &&$n^3\langle\tilde\Delta_2\rangle^{(1)}$ &&$n^3\langle Q\rangle^{(1)}$&&$n^3\langle Q_1\rangle^{(1)}$\\
\hline
  2&   -2&.714\,702(7)      & 0&.836\,043\,833(2)    &0&.990\,15(6)      & -0&.086\,834\,3(5)      &1&.387\,38(2)      & -0&.062\,592\,5(6)    & -0&.979\,5(2)    \\                
  3&   -2&.646\,522(7)      & 0&.956\,097\,734(8)    &0&.956\,64(2)      & -0&.090\,912\,22(4)     &1&.771\,10(1)      & -0&.054\,894\,79(7)   & -0&.947\,65(3)   \\                
  4&   -2&.629\,961(2)      & 1&.012\,181\,008(2)    &0&.943\,236(4)     & -0&.090\,794\,23(8)     &1&.817\,491(5)     & -0&.051\,075\,7(1)    & -0&.933\,920(6)  \\
  5&   -2&.628\,160(2)      & 1&.046\,222\,668(7)    &0&.937\,276(9)     & -0&.090\,257\,1(6)      &1&.806\,723(6)     & -0&.048\,888\,0(9)    & -0&.927\,39(2)   \\
  6&   -2&.630\,7266(6)     & 1&.069\,350\,02(1)     &0&.934\,361(9)     & -0&.089\,747\,2(4)      &1&.784\,751(2)     & -0&.047\,486\,8(4)    & -0&.923\,98(3)   \\
  7&   -2&.634\,471174(7)   & 1&.086\,152\,33(2)     &0&.932\,804(2)     & -0&.089\,319\,421(5)    &1&.762\,15909(3)   & -0&.046\,518\,86(1)   & -0&.921\,958(9)  \\
  8&   -2&.638\,346(8)      & 1&.098\,933\,0301(4)   &0&.931\,95(1)      & -0&.088\,969\,1995(6)   &1&.741\,54(2)      & -0&.045\,810\,6689(2) & -0&.920\,733(9)  \\
  9&   -2&.642\,00(2)       & 1&.108\,989\,541(2)    &0&.931\,51(4)      & -0&.088\,680\,761(1)    &1&.723\,33(5)      & -0&.045\,271\,315(5)  & -0&.919\,99(6)   \\
 10&   -2&.645\,3131(8)     & 1&.117\,112\,384(2)    &0&.931\,252(2)     & -0&.088\,440\,739(4)    &1&.707\,426(2)     & -0&.044\,847\,2687(7) & -0&.919\,466(4)  \\
 11&   -2&.648\,288(1)      & 1&.123\,811\,999(4)    &0&.931\,15(1)      & -0&.088\,238\,73(1)     &1&.693\,501(3)     & -0&.044\,505\,186(7)  & -0&.919\,16(4)   \\
 12&   -2&.650\,926(1)      & 1&.129\,433\,168(3)    &0&.931\,089(3)     & -0&.088\,066\,6111(1)   &1&.681\,299(2)     & -0&.044\,223\,7975(1) & -0&.918\,882(6)  \\
 13&   -2&.653\,2822(5)     & 1&.134\,217\,357(3)    &0&.931\,092(2)     & -0&.087\,918\,5154      &1&.670\,510(1)     & -0&.043\,988\,1411(4) & -0&.918\,715(6)  \\
 14&   -2&.655\,386(5)      & 1&.138\,338\,830(4)    &0&.931\,131(5)     & -0&.087\,789\,85(1)     &1&.660\,919(9)     & -0&.043\,787\,99(1)   & -0&.918\,603(5)  \\
 15&   -2&.657\,28(2)       & 1&.141\,926\,5312      &0&.931\,21(3)      & -0&.087\,677\,10(1)     &1&.652\,32(4)      & -0&.043\,615\,92(2)   & -0&.918\,58(5)   \\
 16&   -2&.658\,97(2)       & 1&.145\,077\,9706      &0&.931\,27(2)      & -0&.087\,577\,508(8)    &1&.644\,62(4)      & -0&.043\,466\,46(2)   & -0&.918\,52(4)   \\
 17&   -2&.660\,4553(9)     & 1&.147\,868\,2073(7)   &0&.931\,306(4)     & -0&.087\,488\,99(1)     &1&.637\,710(2)     & -0&.043\,335\,36(2)   & -0&.918\,43(2)   \\
 18&   -2&.661\,818(5)      & 1&.150\,356\,0370(9)   &0&.931\,37(1)      & -0&.087\,409\,79(5)     &1&.631\,395(9)     & -0&.043\,219\,47(6)   & -0&.918\,40(3)   \\
 19&   -2&.663\,10(2)       & 1&.152\,588\,113(6)    &0&.931\,50(3)      & -0&.087\,338\,2(2)      &1&.625\,54(5)      & -0&.043\,116\,7(3)    & -0&.918\,47(5)   \\
 20&   -2&.664\,1846(2)     & 1&.154\,601\,987(1)    &0&.931\,532736(4)  & -0&.087\,273\,898(8)    &1&.620\,3132(5)    & -0&.043\,024\,11(2)   & -0&.918\,40964(6)\\
 21&   -2&.665\,2014(4)     & 1&.156\,428\,170(1)    &0&.931\,5960(3)    & -0&.087\,215\,287(8)    &1&.615\,4478(8)    & -0&.042\,940\,88(1)   & -0&.918\,3929(3) \\
 22&   -2&.666\,1303(9)     & 1&.158\,091\,742(3)    &0&.931\,662(2)     & -0&.087\,161\,811(3)    &1&.610\,948(2)     & -0&.042\,865\,493(2)  & -0&.918\,392(4)  \\
 23&   -2&.666\,98857(8)    & 1&.159\,613\,509(4)    &0&.931\,7420(3)    & -0&.087\,112\,79(4)     &1&.606\,7565(2)    & -0&.042\,796\,92(7)   & -0&.918\,4225(7) \\
 24&   -2&.667\,765(4)      & 1&.161\,010\,862(4)    &0&.931\,802(4)     & -0&.087\,067\,721(5)    &1&.602\,0(9)       & -0&.042\,734\,294(3)  & -0&.918\,424(7)  \\
 25&   -2&.668\,46(2)       & 1&.162\,298\,49(1)     &0&.931\,85(2)      & -0&.087\,026\,13(3)     &1&.599\,24(5)      & -0&.042\,676\,84(5)   & -0&.918\,41(2)   \\
 26&   -2&.669\,129(4)      & 1&.163\,488\,784(5)    &0&.931\,92(1)      & -0&.086\,987\,66(3)     &1&.598\,(2)        & -0&.042\,623\,90(4)   & -0&.918\,47(4)   \\
 27&   -2&.669\,72(4)       & 1&.164\,592\,423(3)    &0&.931\,90(4)      & -0&.086\,951\,95(2)     &1&.591\,(1)        & -0&.042\,575\,08(3)   & -0&.918\,34(4)   \\
 28&   -2&.670\,3(2)        & 1&.165\,618\,523(3)    &0&.932\,0(2)       & -0&.086\,918\,72(2)     &1&.589\,7(4)       & -0&.042\,529\,82(3)   & -0&.918\,5(3)    \\
 29&   -2&.670\,91(5)       & 1&.166\,574\,9950      &0&.932\,32(3)      & -0&.086\,887\,4154      &1&.586\,59(7)      & -0&.042\,488\,3341(2) & -0&.918\,89(5)   \\
 30&   -2&.671\,8(1)        & 1&.167\,468\,67(5)     &0&.932\,9(2)       & -0&.086\,858\,5(5)      &1&.583\,1(2)       & -0&.042\,449\,0(8)    & -0&.919\,9(5)    \\
 31&   -2&.671\,808(1)      & 1&.168\,305\,550(4)    &0&.932\,26(1)      & -0&.086\,831\,55(5)     &1&.581\,4(2)       & -0&.042\,412\,1(6)   & -0&.918\,62(4)   \\
 32&   -2&.672\,1974(7)     & 1&.169\,090\,888(3)    &0&.932\,347(2)     & -0&.086\,805\,9(5)      &1&.585(6)          & -0&.042\,378\,1(8)    & -0&.918\,759(6)  \\
 33&   -2&.672\,827(3)      & 1&.169\,829\,304(4)    &0&.932\,346(8)     & -0&.086\,781\,82(1)     &1&.576\,8(6)       & -0&.042\,346\,24(2)   & -0&.918\,669(6) \\
 34&   -2&.71(4)            & 1&.170\,525(4)         &0&.96(\,3)         & -0&.086\,758\,0(2)      &1&.577(2)          & -0&.042\,317\,9(3)    & -0&.9(1\,)       \\
 35&   -4&.7(2\,)           & 1&.171\,18(1)          &0&.8(1\,)          & -0&.086\,733\,7(2)      &1&.9(2\,)          & -0&.042\,293\,3(5)    & -0&.8(2\,)       \\
\hline
\hline
\end{tabular}
\end{table*}

\begin{table*}[tb]
\caption{Mass polarization coefficients $X^{(1)}$ scaled by $n^3$ for matrix elements needed to calculate the relativistic and QED contributions to the energy for the $1snp\;^3P$ states of helium. See Eqs.\ (\ref{eq:X(1)}) and (\ref{eq:X(M)}).  Unit are atomic units.}
\label{Table:tripletM}
\begin{tabular}{rr@{}lr@{}lr@{}lr@{}lr@{}lr@{}lr@{}l}
\hline\hline
 $n$  &&$n^3\langle p_1^4\rangle^{(1)}$ &&$n^3\langle B_2\rangle^{(1)}$ &&$n^3\pi\langle\delta(\br_1)\rangle^{(1)}$&&$n^3\pi\langle\delta(\br_{12}\rangle^{(1)}$ &&$n^3\langle\tilde\Delta_2\rangle^{(1)}$ &&$n^3\langle Q\rangle^{(1)}$&&$n^3\langle Q_1\rangle^{(1)}$\\
\hline
  2    &4&.739\,181(5)       &1&.218\,981\,024(2)  & -1&.801\,67(4)        &0&.0      &9&.847\,00(1)        &0&.068\,254\,6814    &  1&.792\,9(1)     \\
  3    &3&.337\,71(3)        &1&.153\,763\,743(4)  & -1&.239\,53(8)        &0&.0      &6&.519\,46(5)        &0&.048\,377\,7674(1) &  1&.254\,9(2)     \\
  4    &2&.858\,481(2)       &1&.142\,859\,050(9)  & -1&.039\,477(2)       &0&.0      &5&.471\,822(5)       &0&.041\,233\,4703(1) &  1&.061\,46(1)    \\
  5    &2&.636\,103(2)       &1&.143\,578\,423(2)  & -0&.943\,980(4)       &0&.0      &5&.020\,532(3)       &0&.037\,668\,945(1)  &  0&.968\,406(8)   \\
  6    &2&.512\,689(5)       &1&.146\,854\,157(6)  & -0&.889\,77(2)        &0&.0      &4&.786\,45(1)        &0&.035\,548\,835(3)  &  0&.915\,36(6)    \\
  7    &2&.435\,983(2)       &1&.150\,478\,209(5)  & -0&.855\,37(1)        &0&.0      &4&.649\,766(3)       &0&.034\,146\,9316(1) &  0&.881\,45(4)    \\
  8    &2&.384\,4467(6)      &1&.153\,865\,7721    & -0&.831\,8497(4)      &0&.0      &4&.563\,109(1)       &0&.033\,152\,3569    &  0&.858\,1717(3)  \\
  9    &2&.347\,82(1)        &1&.156\,883\,288(2)  & -0&.814\,86(3)        &0&.0      &4&.504\,77(3)        &0&.032\,410\,5642    &  0&.841\,29(7)    \\
 10    &2&.320\,7032(2)      &1&.159\,531\,8583    & -0&.802\,1601(6)      &0&.0      &4&.463\,7475(4)      &0&.031\,836\,2482    &  0&.828\,724(2)   \\
 11    &2&.299\,88(2)        &1&.161\,850\,6021    & -0&.792\,24(5)        &0&.0      &4&.433\,69(4)        &0&.031\,378\,542(4)  &  0&.818\,8(1)     \\
 12    &2&.283\,4849(7)      &1&.163\,885\,394(3)  & -0&.784\,344(2)       &0&.0      &4&.411\,075(1)       &0&.031\,005\,2548(1) &  0&.810\,939(5)   \\
 13    &2&.270\,279(5)       &1&.165\,678\,876(3)  & -0&.777\,904(7)       &0&.0      &4&.393\,60(1)        &0&.030\,695\,0310    &  0&.804\,50(1)    \\
 14    &2&.259\,447(6)       &1&.167\,267\,871(4)  & -0&.772\,562(7)       &0&.0      &4&.379\,81(1)        &0&.030\,433\,1514    &  0&.799\,13(1)    \\
 15    &2&.250\,43(2)        &1&.168\,683\,228(5)  & -0&.768\,06(3)        &0&.0      &4&.368\,73(5)        &0&.030\,209\,1424    &  0&.794\,61(6)    \\
 16    &2&.242\,816(2)       &1&.169\,950\,546(6)  & -0&.764\,242(7)       &0&.0      &4&.359\,699(4)       &0&.030\,015\,3498(1) &  0&.790\,77(2)    \\
 17    &2&.236\,330(2)       &1&.171\,090\,974(7)  & -0&.760\,960(4)       &0&.0      &4&.352\,247(4)       &0&.029\,846\,051(1)  &  0&.787\,48(1)    \\
 18    &2&.230\,75(1)        &1&.172\,122\,037(9)  & -0&.758\,10(4)        &0&.0      &4&.346\,03(3)        &0&.029\,696\,880(1)  &  0&.784\,61(8)    \\
 19    &2&.225\,87(4)        &1&.173\,058\,359(1)  & -0&.755\,59(6)        &0&.0      &4&.340\,70(7)        &0&.029\,564\,452(1)  &  0&.782\,1(1)     \\
 20    &2&.221\,5717(1)      &1&.173\,912\,111(1)  & -0&.753\,3075(1)      &0&.0      &4&.336\,09181(4)     &0&.029\,446\,0966    &  0&.779\,729(1)   \\
 21    &2&.217\,81(1)        &1&.174\,693\,548(1)  & -0&.751\,33(3)        &0&.0      &4&.332\,18(3)        &0&.029\,339\,6884(2) &  0&.777\,73(6)    \\
 22    &2&.214\,460(6)       &1&.175\,411\,332(2)  & -0&.749\,54(1)        &0&.0      &4&.328\,73(1)        &0&.029\,243\,5043(1) &  0&.775\,91(2)    \\
 23    &2&.211\,46(1)        &1&.176\,072\,818(2)  & -0&.747\,93(1)        &0&.0      &4&.325\,70(3)        &0&.029\,156\,1395(2) &  0&.774\,28(2)    \\
 24    &2&.208\,78(2)        &1&.176\,684\,297(2)  & -0&.746\,45(3)        &0&.0      &4&.323\,0153(7)      &0&.029\,076\,4336(1) &  0&.772\,74(7)    \\
 25    &2&.206\,36(3)        &1&.177\,251\,163(2)  & -0&.745\,16(2)        &0&.0      &4&.320\,64(7)        &0&.029\,003\,4232    &  0&.771\,46(3)    \\
 26    &2&.204\,203(7)       &1&.177\,778\,085(8)  & -0&.743\,967(5)       &0&.0      &4&.318\,621(9)       &0&.028\,936\,2985    &  0&.770\,24(2)    \\
 27    &2&.202\,193(4)       &1&.178\,269\,089(3)  & -0&.742\,826(9)       &0&.0      &4&.316\,5(5)         &0&.028\,874\,3795(1) &  0&.769\,07(2)    \\
 28    &2&.200\,6(4)         &1&.178\,727\,70(1)   & -0&.742\,0(5)         &0&.0      &4&.315\,2(9)         &0&.028\,817\,072(2)  &  0&.768\,4(8)     \\
 29    &2&.197\,787(7)       &1&.179\,157\,006(3)  & -0&.739\,288(6)       &0&.0      &4&.311\,33(1)        &0&.028\,763\,8998(1) &  0&.764\,140(9)   \\
 30    &2&.198\,(2)          &1&.179\,559\,690(4)  & -0&.740\,(2)          &0&.0      &4&.311(2)            &0&.028\,714\,401(9)  &  0&.765\,(3)      \\
 31    &2&.195\,617(4)       &1&.179\,938\,158(4)  & -0&.738\,802(2)       &0&.0      &4&.309\,8239(5)      &0&.028\,668\,2335(2) &  0&.764\,681(3)   \\
 32    &2&.17(2)             &1&.180\,294\,51(6)   & -0&.736\,954(5)       &0&.0      &4&.307\,313(3)       &0&.028\,625\,0664    &  0&.761\,84(1)    \\
 33    &2&.38(2)             &1&.180\,630\,63(1)   & -0&.737\,044(1)       &0&.0      &4&.307\,312(9)       &0&.028\,584\,6082    &  0&.762\,46(1)   \\
 34    &2&.1(1)              &1&.180\,948(4)       & -0&.778(3)            &0&.0      &4&.359\,(8)          &0&.028\,546\,2(4)    &  0&.841(7)      \\
 35   &&--                   &1&.181\,255(5)       & -1&.0(5)              &0&.0      &5&.(1)               &0&.028\,511\,1(2)    &  1&.3(6)        \\
\hline
\hline
\end{tabular}
\end{table*}

Matrix elements of the various quantities needed to calculate the relativistic and QED contributions to the energy are listed in Table \ref{Table:singlet0} for the $1snp\;^1P$ states and Table \ref{Table:triplet0} for the $1snp\;^3P$ states.  All the matrix elements are scaled by a factor of $n^3$, both to avoid tabulating small numbers for large $n$, and to show that the scaled matrix elements tend asymptotically to a constant.  For the cases of $\langle p_1^4\rangle$, $\pi\langle\delta(\br_1)\rangle$ and $\langle Q_1\rangle$, the $n$-independent contribution from the inner $1s$ electron is first subtracted (i.e.\ 40, 4, and $\frac{16}{\pi}\ln2$ respectively) so that only the residual remainder, scaled by $n^3$, is tabulated. It is only the residual remainder that contributes to the ionization energy.  

Tables \ref{Table:singletM} and \ref{Table:tripletM} list the first- (plus higher-) order correction coefficients $X^{(1)}$ $(X= p_1^4, \cdots)$ multiplying $\mu/M$ due to the mass polarization term $-(\mu/M)\nabla_1\bdot\nabla_2$ in the Hamiltonian. The coefficients are obtained by finite differencing according to the formula
\begin{equation}
\label{eq:X(1)}
X^{(1)} = \frac{X(\mu/M) - X(0)}{\mu/M}\,,
\end{equation}
where $X(\mu/M)$ and $X(0)$ are the matrix elements calculated with and without the mass polarization term included in the Hamiltonian, and so
they sum to infinity the perturbation series at the current CODATA value for $\mu/M$.     
Since the future CODATA value may change, 
the value for any other $X(\mu/M')$ close to $X(\mu/M)$ is then very well approximated by the linearized expression
\begin{equation}
\label{eq:X(M)}
X(\mu/M') = X(0) + \frac{\mu}{M'}X^{(1)}
\end{equation}
In Tables \ref{Table:singlet0} to \ref{Table:tripletM} and those to follow, it is the quantities $X(0)$ and $X^{(1)}$ that are tabulated.  Equation
(\ref{eq:X(M)}) is also useful in calculating isotope shifts, up to terms or order $(\mu/M)^2$.

Our previous work \cite{Bondy2025,Drake2026} provided detailed examples of the calculated contributions to the ionization energy of the $27\,^1P_1$ and $27\,^3P_J$ states of $^4$He relative to $^4$He$^+(1s)$. Detailed tabulations for all $P$-states up to $n = 35$ are available in Ref.\ \cite{data}. The dominant source of uncertainty ($\pm$1 kHz) is the residual QED contribution of order $\alpha^4$ Ry.  However, this level of theoretical uncertainty is already much better than for the low-lying states of helium, and it establishes absolute points of reference for transitions to low-lying states. Our previous Ref.\ \cite{Drake2026} tabulates the calculated ionization energies for all the $n\,^1P_1$ and $n\,^3P_{\rm c}$ (centroid) states of helium for 
$2\le n \le 35$, together with the fine-structure energy shifts relative to the centroid energy.

\section{Quantum Defects and $1/n$ Expansions}
\subsection{Nonrelativistic Energies}

The main purpose of the present work is to obtain extrapolations to higher $n$ by means of the quantum defect method \cite{Edlen,Seaton,Jungen}.  However, the high precision now available provides an opportunity to test the validity of the method itself.
According to the usual formulation of QDT, the term energies for a
quasi-hydrogenic atom with effective nuclear charge $Z_{\rm eff} = 1$ are given by
\begin{equation}
E_n = -\frac{R_M}{n*^2}
\end{equation}
where $R_M = \frac{M}{m_e+M}R_\infty$ is the reduced mass Rydberg for an atom of mass $M$ (to be discussed further), 
 $n* = n - \delta(n*)$ is the effective principal quantum number,
and $\delta(n*)$ is the quantum defect defined by the Ritz expansion
\begin{equation}
\label{eq:deltan}
\delta(n*) = \delta_0 + \frac{\delta_2}{(n-\delta)^2} + \frac{\delta_4}{(n-\delta)^4} + \cdots
\end{equation}]   
containing only even powers.  The question to be addressed is: under what conditions is the quantum defect expansion exact, 
especially concerning the precise value used for $R_M$ and the Ritz expansion.  Hartree's proof \cite{Hartree,Langer} guarantees that, for
for a Hamiltonian $ H = H_C + \lambda V$
where $H_C$ is a purely Coulombic Hamiltonian and $V$ is a local short-range spherically symmetric
correction potential, then the eigenvalues are given exactly by the Ritz expansion for the quantum defect.
One might define a ``Ritz defect" \cite{Advances} to be a deviation from Hartree's proof.  Various extensions to Hartree's proof involving long-range
potentials have been discussed in the literature (see for example Refs.\ \cite{Seaton1966,Seaton,Advances,Jungen,SwainsonDrake}, but no Ritz defect has been found,
at least in the nonrelativistic infinite-nuclear-mass limit.  Relativistic versions of QDT have been developed {\cite{Johnson77,Jacobs2022}, but the
coefficients in a $1/n$ expansion do not meet the requirements for the Ritz expansion to be valid; i.e.\ for the odd $\delta_i$ to vanish. 

\begin{table}[tb]
\caption{Quantum defect parameters to fit the nonrelativistic energies for the  $1snp\;^1P$ and 
$1snp\;^3P$ states of helium, including both the infinite nuclear mass and finite nuclear mass cases.
For the finite mass case, see Eq.\ \ref{eq:E_ncorrected}.}
\label{Table:QDfit}
\begin{tabular}{rr@{}lr@{}l}
\hline\hline
 $t$  &&  $\delta_t(^1P)\times10^6$  && $\delta_t(^3P)\times10^6$  \\
\hline
\multicolumn{5}{c}{Infinite nuclear mass case}\\
 0 &   -12\,114&.192\,143\,935\,069\,93(38)&     68\,293&.614\,172\,491\,0231(10)\\
 2 &       7507&.893\,176\,176\,17(32)     &    -18\,636&.050\,617\,7460(14)     \\
 4 &    13\,958&.924\,048\,95(14)          &    -12\,317&.146\,265\,87(42)       \\
 6 &       4880&.202\,286(16)              &       -8064&.197\,353(37)           \\
 8 &        889&.569\,21(57)               &       -4612&.819\,8(12)             \\
 10&        700&.934\,3(87)                &       -1423&.315(18)                \\
 12&        159&.457(56)                   &        -743&.77(11)                 \\
 14&       -248&.55(12)                    &        -518&.31(21)                 \\
                                                                                 \\
 \multicolumn{5}{c}{Finite nuclear mass case for $^4$He}\\
 0 &   -12\,170&.559\,891\,525\,778(10)    &     68\,355&.695\,242\,244(65)      \\
 2 &     7\,523&.086\,479\,853\,9(28)      &    -18\,620&.076\,142(22)           \\
 4 &    13\,981&.956\,941\,91(21)          &    -12\,331&.288\,9(23)             \\
 6 &     4\,891&.575\,670\,98(21)          &       -8083&.805(83)                \\
 8 &        893&.399\,450(80)              &       -4618&.8(1.2)                 \\
 10&        713&.071\,18(85)               &       -1452&.5(7.6)                 \\
 12&         57&.953\,8(22)                &        -889&.8(15.2)                \\
\hline\hline                              
\end{tabular}
\end{table}

For these reasons, our strategy is to use the Ritz QDT only for the nonrelativistic energies where it is expected to be valid.
Starting with the infinite nuclear mass case, the results of the quantum defect fit are shown in the upper half of Table \ref{Table:QDfit}.
Equation \ref{eq:deltan} is solved iteratively, starting with $\delta = 0$, and adding term by term until the $\delta_i$ start increasing in size.
The uncertainty is determined by a ``bootstrap" method.  The accuracy of the fit is sufficient to reproduce nearly the full 20-figure accuracy of
the input energies from Table \ref{Table:nonrelE}, and in fact provides a compact and highly accurate representation of the entire Rydberg series from $n = 2$ to 35.  The results confirm the validity of the Ritz expansion for the nonrelativistic energy to an unprecedented 20-figure accuracy.

\begin{table*}[tb]
\caption{First-order ($-\langle\nabla_1\bdot\nabla_2\rangle$) and second-order ($\langle\nabla_1\bdot\nabla_2\rangle^{(2)}$) mass polarization matrix
elements for the $1snp\;^1P$ and $1snp\;^3P$ states of helium with the assumed value
$\mu/M = 1.370\,745\,634\,614\times10^{-4}$ for the ratio of the reduced electron mass to the $^4$He nuclear mass.  The second-order
term is obtained by finite differencing, and so it includes higher-order contributions.  Units are atomic units.}   
\label{Table:mass_pol}
\begin{tabular}{rr@{}lr@{}lr@{}lr@{}l}
\hline\hline
    & \multicolumn{4}{c}{$1snp\;^1P$}                   &\multicolumn{4}{c}{$1snp\;^3P$} \\          
 $n$ &&   $-\langle\nabla_1\bdot\nabla_2\rangle$ && $\langle\nabla_1\bdot\nabla_2\rangle^{(2)}$&& $-\langle\nabla_1\bdot\nabla_2\rangle$&& $\langle\nabla_1\bdot\nabla_2\rangle^{(2)}$\\
 \hline        
  2&  0&.368\,356\,199\,427(1)&$-0$&.346\,170\,586(9)   &    $-0$&.516\,579\,400\,215   &$-0$&.639\,679\,1821(4)   \\  
  3&  0&.392\,797\,271\,608   &$-0$&.283\,292\,7158(1)  &    $-0$&.495\,963\,044\,163   &$-0$&.397\,903\,3519(3)   \\  
  4&  0&.400\,315\,108\,408   &$-0$&.250\,247\,64089(5) &    $-0$&.483\,531\,454\,265   &$-0$&.312\,315\,44726(1)  \\  
  5&  0&.403\,752\,721\,417   &$-0$&.231\,316\,7977(1)  &    $-0$&.476\,363\,885\,336   &$-0$&.271\,077\,842941(7) \\  
  6&  0&.405\,660\,644\,442   &$-0$&.219\,275\,9732(5)  &    $-0$&.471\,818\,840\,507(1)&$-0$&.247\,317\,112(5)    \\  
  7&  0&.406\,850\,237\,036   &$-0$&.211\,010\,77865(7) &    $-0$&.468\,709\,771\,116   &$-0$&.232\,022\,21(3)     \\  
  8&  0&.407\,652\,183\,495   &$-0$&.205\,012\,661(1)   &    $-0$&.466\,459\,663\,765   &$-0$&.221\,413\,598(2)    \\  
  9&  0&.408\,224\,048\,777   &$-0$&.200\,473\,383(1)   &    $-0$&.464\,760\,279\,013   &$-0$&.213\,650\,983(2)    \\  
 10&  0&.408\,649\,517\,367   &$-0$&.196\,924\,5664(7)  &    $-0$&.463\,433\,633\,758   &$-0$&.207\,738\,129(1)    \\  
 11&  0&.408\,976\,748\,185   &$-0$&.194\,077\,21(4)    &    $-0$&.462\,370\,338\,806   &$-0$&.203\,091\,52984(1)  \\  
 12&  0&.409\,235\,226\,533   &$-0$&.191\,743\,978(1)   &    $-0$&.461\,499\,685\,423   &$-0$&.199\,347\,923(1)    \\  
 13&  0&.409\,443\,918\,117   &$-0$&.189\,798\,2879(2)  &    $-0$&.460\,774\,038\,341   &$-0$&.196\,269\,9813(2)   \\  
 14&  0&.409\,615\,524\,022   &$-0$&.188\,151\,766797(1)&    $-0$&.460\,160\,183\,157   &$-0$&.193\,696\,1289(6)   \\  
 15&  0&.409\,758\,840\,432(1)&$-0$&.186\,740\,735(9)   &    $-0$&.459\,634\,278\,5(1)  &$-0$&.191\,513\,1(9)      \\  
 16&  0&.409\,880\,134\,8(2)  &$-0$&.185\,519\,(2)      &    $-0$&.459\,178\,781\,79(8) &$-0$&.189\,637\,9(6)      \\  
 17&  0&.409\,983\,982\,84(3) &$-0$&.184\,449\,6(2)     &    $-0$&.458\,780\,505\,040(4)&$-0$&.188\,012\,54(3)     \\  
 18&  0&.410\,073\,796\,87(5) &$-0$&.183\,507\,6(3)     &    $-0$&.458\,429\,348\,638(6)&$-0$&.186\,587\,99(5)     \\  
 19&  0&.410\,152\,168\,74(5) &$-0$&.182\,670\,9(4)     &    $-0$&.458\,117\,449\,033(3)&$-0$&.185\,330\,26(2)     \\  
 20&  0&.410\,221\,099\,68(1) &$-0$&.181\,921\,91(7)    &    $-0$&.457\,838\,594\,6(1)  &$-0$&.184\,212\,6(7)      \\  
 21&  0&.410\,282\,157\,863   &$-0$&.181\,248\,83635(7) &    $-0$&.457\,587\,815\,049   &$-0$&.183\,211\,621048(7) \\  
 22&  0&.410\,336\,587\,83(2) &$-0$&.180\,641\,0(2)     &    $-0$&.457\,361\,086\,08(2) &$-0$&.182\,311\,4(1)      \\  
 23&  0&.410\,385\,388\,978(1)&$-0$&.180\,089\,021(4)   &    $-0$&.457\,155\,115\,70(1) &$-0$&.181\,497\,68(8)     \\  
 24&  0&.410\,429\,372\,27(8) &$-0$&.179\,583\,9(5)     &    $-0$&.456\,967\,185\,313   &$-0$&.180\,757\,234(1)    \\  
 25&  0&.410\,469\,202\,83(4) &$-0$&.179\,121\,7(3)     &    $-0$&.456\,795\,029\,965(4)&$-0$&.180\,080\,25(3)     \\  
 26&  0&.410\,505\,430\,27(1) &$-0$&.178\,697\,38(8)    &    $-0$&.456\,636\,747\,588(3)&$-0$&.179\,460\,61(2)     \\  
 27&  0&.410\,538\,512\,66(1) &$-0$&.178\,305\,65(8)    &    $-0$&.456\,490\,729\,790   &$-0$&.178\,890\,9282(1)   \\  
 28&  0&.410\,568\,834\,70(9) &$-0$&.177\,943\,4(7)     &    $-0$&.456\,355\,607\,079(9)&$-0$&.178\,365\,20(6)     \\  
 29&  0&.410\,596\,721\,82(9) &$-0$&.177\,608\,983(1)   &    $-0$&.456\,230\,206\,(1)   &$-0$&.177\,882\,0(1)      \\  
 30&  0&.410\,622\,450\,3(1)  &$-0$&.177\,294\,53(4)    &    $-0$&.456\,113\,515\,86(3) &$-0$&.177\,427\,3(1)      \\  
 31&  0&.410\,646\,257\,45(4) &$-0$&.177\,002\,2(3)     &    $-0$&.456\,004\,660\,93(3) &$-0$&.177\,006\,0(2)      \\  
 32&  0&.410\,668\,347\,240(2)&$-0$&.176\,730\,3099(2)  &    $-0$&.455\,902\,878\,9(2)  &$-0$&.176\,614\,243(7)    \\  
 33&  0&.410\,688\,895\,68(7) &$-0$&.176\,474\,37(1)    &    $-0$&.455\,807\,502\,74(2) &$-0$&.176\,246\,89(7)     \\  
 34&  0&.410\,708\,056\,23(4) &$-0$&.176(8)             &    $-0$&.455\,717\,947\,51(9) &$-0$&.176(8)              \\  
 35&  0&.410\,725\,962\,9(7)  &$-0$&.169(6)             &    $-0$&.455\,633\,696\,2(6)  &$-0$&.17(1)               \\  
\hline
\hline
\end{tabular}
\end{table*} 

\begin{table}[tb]              
\caption{$1/n^t$ expansion expansion coefficients for the first- and second-order
mass polarization contribution to the energy for the $1snp\;^1P$ and $1snp\;^3P$ states of helium. Units are atomic units.}                     
\label{Table:mass_pol_exp}      
\begin{tabular}{rr@{}lr@{}l}   
\hline\hline                   
 $t$ && $c(\bp_1\bdot\bp_2)_t^{(1)}$&&$c(\bp_1\bdot\bp_2)_t^{(2)}$\\
\hline 
\multicolumn{5}{c}{$1snp\;^1P$}\\
 2 &  0&.0             &    -0&.500000       \\
 3 &  0&.411\,2423(26) &    -0&.168\,636(14) \\
 4 & -0&.014\,92(11)   &    -0&.2474(7)      \\
 5 & -0&.1109(19)      &    -0&.378(12)      \\
 6 &  0&.027(15)       &     0&.23(10)       \\
 7 & -0&.19(6)         &     0&.0(4)         \\
 8 &  0&.11(11)        &     0&.2(8)         \\
 \\
 \multicolumn{5}{c}{$1snp\;^3P$}\\
 2 &  0&.0             &    -0&.50000        \\
 3 & -0&.452\,876(8)   &    -0&.165\,32(8)   \\
 4 & -0&.0929(4)       &    -0&.335(4)       \\
 5 & -0&.127(7)        &    -0&.83(6)        \\
 6 & -0&.02(6)         &    -0&.7(5)         \\
 7 &  0&.23(23)        &     0&.1(1.7)       \\
 8 & -0&.3(5)          &    -3&.(3)          \\
\hline\hline  
\end{tabular} 
\end{table}   
             
The finite nuclear mass case requires further discussion due to the mass polarization term in the Hamiltonian, which arises physically from the motion of the nucleus in the center-of-mass (CM) frame.  Table \ref{Table:mass_pol} lists the first- and second-order matrix elements of the mass polarization operator $-\nabla_1\bdot\nabla_2$.  The first-order term is calculated directly as a matrix element of the infinite mass wave function, and the second-order term is estimated according to the finite-difference formula
\begin{equation}
\langle\nabla_1\bdot\nabla_2\rangle^{(2)} \simeq \frac{E(\mu/M) - E(0) + (\mu/M)\langle\nabla_1\bdot\nabla_2\rangle}{(\mu/M)^2}
\end{equation}
 It therefore approximates the second-order term and sums to infinity the perturbation series in $\mu/M$ at the assumed CODATA value for $\mu/M$.  
 
To further reveal the significance of these matrix elements in terms of QDT, Table \ref{Table:mass_pol_exp} lists the coefficients in a $1/n$ expansion
fit to the data in Table \ref{Table:mass_pol}. The first-order terms decrease in proportion to $1/n^3$ as expected, but the second-order terms have a 
leading $-2/n^2$ dependence that is state-independent.  It is therefore essential to subtract it from the input energies for the finite mass case 
before performing a QDT fit.  The corrected fitting formula is thus
\begin{equation}
\label{eq:E_ncorrected}
E_n = -\frac{R_M}{n*^2} -\frac{R_M}{n^2}\left(\frac{\mu}{M}\right)^2
\end{equation}
The same $-(R_M/n^2)(\mu/M)^2$ was derived previously in the asymptotic limit of high-$L$ Rydberg states by use of Jacobi coordinates \cite{Drachman86,DrakeYan92}, and included in previous discussions of QDT \cite{Advances}.  The results in Table \ref{Table:mass_pol} confirm the asymptotic result for low-$L$ $P$-states with the same asymptotic coefficient. This is the form of the quantum defect expansion used to obtain the $\delta_i$ parameters as listed in the bottom half of Table \ref{Table:QDfit}.  A further relativistic correction is discussed in the following section.   

Alternatively, if the quantum defect is small, the value of the reduced mass Rydberg can be adjusted to take the $-(R_M/n^2)(\mu/M)^2$ term into account with an expression of the form 
\begin{equation}
\label{eq:En+}
E_n = -R_{M}^{(+)}\frac{1}{n*^2}
\end{equation}
with $R_{M}^{(+)}$ defined to be
\begin{equation}
R_{M}^{(+)} = R_\infty\frac{M+m_e}{M+2m_e}
\end{equation}
in place of $R_M=R_\infty\frac{M}{M+m_e}$. Physically, this is the Rydberg for an electron outside a He$^+$ core instead of a bare He$^{++}$ nucleus. This provides a theoretical justification for what is commonly done on phenomenological grounds in quantum defect fits to data.
Note that
\begin{equation}
\frac{R_M^{(+)}}{R_M} = \frac{1}{1-(\mu/M)^2} = 1 + \left(\frac{\mu}{M}\right)^2 + \left(\frac{\mu}{M}\right)^4 + \cdots 
\end{equation}
and so Eqs.\ (\ref{eq:E_ncorrected}) and (\ref{eq:En+}) are equivalent to order $(\mu/M)^2$ with an error term of order $(\mu/M)^4$.

\begin{table*}[tb]
\caption{$1/n^t$ expansion coefficients $c(X)_t$ for the spin-independent matrix elements needed to calculate the relativistic and QED contributions to the energy for the $1snp\;^1P$ and $1snp\;3P$ states of helium, and corresponding mass polarization coefficients $c(X)_t^{(1)}$. Units are atomic units.}
\label{1/nSI}
\begin{tabular}{rr@{}lr@{}lr@{}lr@{}lr@{}lr@{}lr@{}l}
\hline\hline
 $t$  &&$c(p_1^4)_t$ &&$c(H_2)_t/alpha^2$      && $c(\delta(\br_1))_t$  &&$c(\delta(\br_{12}))_t$ &&$c(\tilde\Delta_{\rm 2,res})_t$ &&$c(Q)_t$ &&$c(Q_1)_t$\\
\hline
\multicolumn{15}{c}{$1snp\;^1\!P$ matrix element expansions}\\
  0 & 40&.000000     &    0&.000000      &   4&.000000      &   0&.000000       &  0&.000000   &      0&.000000      &    &$\frac{16}{\pi}\ln2$\\
 -3 & -0&.437302(12) &   -0&.1950586(14) &   0&.01072794(14)&   0&.007599141(3) & -3&.51472(7) &      0&.027129179(8)&  -7&.06033917(16) \\
 -4 &  0&.3910(5)    &    0&.00712(6)    &  -0&.000390(8)   &  -0&.00027613(15) &  2&.1286(25) &     -0&.0009861(5)  &  -0&.000003(11)   \\
 -5 & -0&.024(6)     &    0&.1023(9)     &   0&.00178(15)   &  -0&.006460(3)    &  0&.47(3)    &      0&.000767(10)  &   0&.00009(29)    \\
 -6 &  0&.01(4)      &   -0&.007(6)      &   0&.0002(14)    &   0&.00069(4)     & -0&.04(16)   &      0&.00093(11)   &  -0&.008(4)       \\
 -7 &  0&.12(9)      &    0&.051(16)     &  -0&.027(7)      &  -0&.00108(19)    &  0&.5(4)     &     -0&.0006(6)     &   0&.013(22)      \\
 -8 &  0&.15(7)      &    0&.032(15)     &   0&.005(14)     &   0&.0010(5)      &  0&.51(29)   &      0&.0016(15)    &  -0&.06(6)        \\
 \multicolumn{15}{c}{$\mu/M$ mass polarization coefficient $c(X)_t^{(1)}$, $X = p_1^4, \cdots$}      \\
 -3 & -2&.6901(15)   &    1&.193803(20)  &   0&.2972(7)     &  -0&.027375(4)    &  1&.509(5)   &     -0&.0413522(27) &  -0&.9191(21)     \\
 -4 &  0&.60(5)      &   -0&.8024(12)    &  -0&.021(17)     &  -0&.00832(22)    &  2&.44(14)   &     -0&.03181(15)   &   0&.03(5)        \\
 -5 & -1&.5(3)       &    0&.379(25)     &   0&.13(7)       &   0&.003(3)       & -4&.4(1.0)   &     -0&.0339(26)    &  -0&.36(18)       \\
 -6 &  0&.4(5)       &   -0&.19(19)      &    &             &   0&.023(13)      & -1&.9(1.6)   &      0&.029(19)     &    &              \\
 -7 &   &            &   -0&.44(28)      &    &             &    &              &   &          &     -0&.04(6)       &    &              \\
 -8 &   &            &     &             &    &             &    &              &   &          &      0&.07(6)       &    &              \\
\multicolumn{15}{c}{$1snp\;^3\!P$ matrix element expansions}                                                                             \\
  0 & 40&.000000     &    0&.000000      &   4&.000000      &   0&.000000       &  0&.000000   &      0&.000000      &    &$\frac{16}{\pi}\ln2$\\
 -3 &  0&.404605(6)  &    0&.2625288(17) &  -0&.10271175(29)&    &              &  0&.683366(20)&     0&.02502589(9) &  -7&.0603388(6)   \\
 -4 &  0&.4580(3)    &    0&.05380(9)    &  -0&.021048(20)  &    &              &  2&.1403(12)  &     0&.005129(5)   &  -0&.00003(3)     \\
 -5 &  0&.311(6)     &    0&.0083(16)    &  -0&.0167(5)     &    &              &  0&.939(23)   &     0&.01085(10)   &   0&.0007(7)      \\
 -6 &  0&.07(6)      &   -0&.018(14)     &   0&.003(5)      &    &              &  0&.24(21)    &     0&.0016(8)     &   0&.328(6)       \\
 -7 & -0&.23(27)     &   -0&.11(6)       &   0&.023(24)     &    &              & -1&.(1)       &    -0&.000(4)      &   0&.113(21)      \\
 -8 &  0&.4(6)       &    0&.05(13)      &  -0&.03(5)       &    &              &  1&.1(2.1)    &     0&.006(7)      &  -0&.044(28)      \\
 \multicolumn{15}{c}{$\mu/M$ mass polarization coefficient $c(X)_t^{(1)}$, $X = p_1^4, \cdots$}                                                                           \\
 -3 &  2&.157(6)     &    1&.1921(5)     &  -0&.229(4)      &    &              &  4&.295(20)   &     0&.0273384(10) &   0&.7431(22)     \\
 -4 &  0&.95(17)     &   -0&.401(13)     &  -0&.16(10)      &    &              &  0&.0(6)      &     0&.03963(6)    &   0&.61(7)        \\
 -5 &  6&.4(1.3)     &    0&.70(12)      &  -1&.0(5)        &    &              & 15&.(5)       &     0&.0477(13)    &   2&.3(5)         \\
 -6 &  4&.0(2.0)     &    0&.4(4)        &    &             &    &              & 14&.(7)       &     0&.056(13)     &   1&.5(1.1)       \\
 -7 &   &            &     &             &    &             &    &              &   &           &     0&.00(6)       &    &              \\
 -8 &   &            &     &             &    &             &    &              &   &           &     0&.16(14)      &    &              \\
\hline
\hline   
\end{tabular}
\end{table*}

\begin{table}[tb]
\caption{$1/n^t$ expansion coefficients $c(X)_t$ for the spin-dependent matrix elements needed to calculate the relativistic contributions to the energy for the $1snp\;^3P$ states of helium, and corresponding $\mu/M$ mass polarization coefficients $c(X)_t^{(1)}$. Units are $\alpha^2$ atomic units.}
\label{1/nSD}
\begin{tabular}{rr@{}lr@{}lr@{}lr@{}l}
\hline\hline
 $t$  && $c(H_{\rm so})$  && $c(H_{\rm soo})$  && $c(H_{\rm ss})$ &&    $c(\tilde\Delta_3)$ \\
\hline
\multicolumn{9}{c}{$1snp\;^3\!P$ matrix element expansions}\\
3 &  1&.5256025(3) &   -2&.200288(13) &   -0&.894235(4) &    -4&.465063(12)\\
4 &  0&.312563(21) &   -0&.4509(6)    &   -0&.18326(21) &    -0&.9149(8)   \\
5 &  0&.0523(5)    &   -0&.342(10)    &   -0&.364(4)    &    -0&.129(17)   \\
6 & -0&.136(6)     &    0&.08(8)      &   -0&.05(4)     &     0&.35(18)    \\
7 & -0&.08(4)      &    0&.4(3)       &    0&.02(15)    &     1&.5(9)      \\
8 & -0&.24(16)     &   -0&.2(7)       &   -0&.2(3)      &    -0&.3(2.0)    \\
\multicolumn{9}{c}{$\mu/M$ mass polarization coefficient $c(X)_t^{(1)}$, $X = H_{\rm so}, \cdots$}                 \\
3 &  3&.0663(17)   &   -3&.9158(17)   &   -1&.11972(17) &    -5&.684(10)   \\
4 &  2&.66(5)      &   -3&.78(5)      &   -1&.454(6)    &    -7&.0(3)      \\
5 &  4&.3(4)       &   -5&.6(5)       &   -1&.63(7)     &   -15&.4(2.5)    \\
6 &  1&.0(7)       &   -3&.8(1.2)     &   -2&.64(25)    &    -5&.(5)       \\
7 &   &            &     &            &   -0&.17(27)    &      &           \\
 \\
\multicolumn{9}{c}{$1snp\;^3\!P-^1\!P$ off-diagonal matrix element expansions}\\ 
3 &  0&.8291725(9)  &   0&.2721294(3)  &     &           &    0&.5442588(29)\\
4 &  0&.06986(5)    &   0&.022925(16)  &     &           &    0&.04585(11)  \\
5 & -0&.0227(9)     &   0&.0919(3)     &     &           &    0&.1838(17)   \\
6 &  0&.018(8)      &   0&.0258(29)    &     &           &   -0&.021(11)    \\
7 & -0&.06(3)       &  -0&.007(13)     &     &           &   -0&.02(3)      \\
8 &  0&.10(7)       &   0&.049(28)     &     &           &    0&.071(28)    \\
\multicolumn{9}{c}{$\mu/M$ mass polarization coefficient $c(X)_t^{(1)}$, $X = H_{\rm so}, \cdots$}                 \\
 & -0&.675(8)     &   -0&.290805(3)  &     &           &    -0&.5820(18)  \\
 & -0&.10(14)     &   -0&.00795(17)  &     &           &    -0&.00(3)     \\
 &  0&.9(5)       &   -0&.119(3)     &     &           &    -0&.34(17)    \\
 &   &            &    0&.467(24)    &     &           &     0&.45(22)    \\
 &   &            &    0&.15(7)      &     &           &      &           \\
\hline
\hline
\end{tabular}
\end{table} 

\subsection{Relativistic and QED Corrections}
Similarly, $1/n$ expansions can be obtained for all the other matrix elements required to calculate the relativistic and QED corrections.  The results are listed in Table \ref{1/nSI} for the spin-independent terms and Table \ref{1/nSD} for the spin-dependent terms. Each matrix element is expressed in the form
\begin{equation}
\langle X\rangle = c(X)_0 + \sum_{t=3}^8\left[ c(X)_t^{(0)} + \frac{\mu}{M}c(X)_t^{(1)}\right]/n^t
\end{equation}
for each operator $X = p_1^4, H_2,$ etc.,  where $c(X)_t^{(0)}$ is the coefficient of $1/n^t$ for infinite nuclear mass, and 
$c(X)_t^{(1)}$ is the finite mass correction coefficient obtained by finite differencing for the case of $^4$He. They therefore sum to infinity the perturbation series in $\mu/M$, at the assumed value for $^4$He, and provide an excellent approximation for slightly adjusted values. The leading
$c(X)_0$ term is zero except for the one-electron operators $p_1^4$, $\delta(\br_1)$, and $Q_1$.  The leading terms of course cancel from the ionization energy with the corresponding hydrogenic He$^+$ terms at the series limit, and so can be dropped.

One exception to the above $\mu/M$ dependence is the relativistic recoil term $\tilde\Delta_2$.  It is of order $m_e/M$ in lowest order, but the $\mu/M$ correction has a $1/n^2$ dependence that is state-independent with the value $8/3$.  The complete matrix element can therefore be written in the form
\begin{equation}          
(m_e/M)\tilde\Delta_2 = (m_e/M)[\tilde\Delta_2^{(0)} + (\mu/M)\tilde\Delta_2^{(1)}]
\end{equation}
with
\begin{equation}
\tilde\Delta_2^{(1)}= \alpha^2\left(
\frac{8}{3n^2}\right) + \tilde\Delta_{\rm 2,res}^{(1)}
\end{equation}
where $\tilde\Delta_{\rm 2,res}^{(1)}$ is the residual part multiplied by $(\mu m_e/M^2)$ that decreases as $1/n^3$ or faster. It is the $1/n$ coefficients of residual part  $\tilde\Delta_{\rm 2,res}^{(1)}$  that are tabulated in Table \ref{1/nSI}.

In addition to the above, it is known from previous work on asymptotic expansions \cite{DrakeYan92,Advances} that the mass polarization correction to the $B_1 + B_4$ terms in the Breit interaction have a leading $1/n^2$ dependence.  With the definition
\begin{eqnarray}
\frac{\mu}{M}\langle B_1^X + B_4^X\rangle &=& \frac{\mu}{M}\left[\langle B_1^X + B_4^X\rangle^{(0)} \right.\nonumber\\
&&\left.\mbox{}+ \frac{\mu}{M}\langle B_1^X + B_4^X\rangle^{(1)}\right]
\end{eqnarray}
then, according to Eq.\ (86) of Ref.\ \cite{DrakeYan92},    
 \begin{equation}
\langle B_1^X + B_4^X\rangle^{(1)} = \case{5}{12}\alpha^2\left(\frac{\mu}{M}\right)^2 + O(\alpha^2/n^3)
\end{equation}
For $Z=2$, and with the approximation $\mu \simeq m_e$, the two contributions combine to give a total of $(\frac83 -\frac53)\alpha^2(\mu/M)^2=\alpha^2(\mu/M)^2$, and so
the total recoil term is thus (in Rydbergs)
\begin{equation}
E_n^{(2)} = -\frac{R_M}{n^2}\left(\frac{\mu}{M}\right)^2(1 - 2\alpha^2)
\end{equation}
The quantum defect formula (\ref{eq:E_ncorrected}) should correspondingly be modified to read
\begin{equation}
\label{eq:E_ncorrected2}
E_n = -\frac{R_M}{n*^2} -\frac{R_M}{n^2}\left(\frac{\mu}{M}\right)^2(1-2\alpha^2)
\end{equation}
If relativistic effects are included, or for experimental data, $E_n^{(2)}$ is the quantity that should be subtracted from (negative) binding energies before performing a quantum defect fit, and then added back in at the end. Otherwise, the nonrelativistic Eq.\ (\ref{eq:E_ncorrected}) should be used, or alternatively Eq.\ (\ref{eq:En+}) with the adjusted $R_M^{(+)}$
Rydberg.
 
\begin{table*}[tb]
\caption{
 Ionization energy of the $1s2s\;^1S_0$ state of $^4$He with 
 $I_{\rm tot}(2\;^1S_0) = \nu_{\rm exp}(2\;^1S_0-n\;^1P_1) + I_{\rm theo}(n\;^1P_1)$.
 The last column gives $\tilde{I}_{\rm tot}(2\;^3S_1)$ obtained by adding the $2\;^1S_0 - 2\;^3S_1$
 transition frequency 192\,510\,702.148\,649 MHz \cite{Rengelink2018,Steinebach2026}.}
 \label{table:results_singlet} 
 \begin{tabular}{r llll}
\hline
\hline
\multicolumn{1}{c}{$n$}& \multicolumn{1}{c}{$\nu_{\rm exp}(2\;^1S_0-n\;^1P_1)^{\rm\,a}$}  &
\multicolumn{1}{c}{$I_{\rm theo}(n\;^1P_1)$ } 
&\multicolumn{1}{c}{$I_{\rm tot}(2\;^1S_0)$}
&\multicolumn{1}{c}{$\tilde{I}_{\rm tot}(2\;^3S_1)$}\\
\hline
  24 &  954\,627\,060.151(30) &    5\,704\,980.3477(16) &  960\,332\,040.499(30)           &   1152842742.647(22) \\
  25 &  955\,074\,118.614(26) &    5\,257\,921.9483(14) &  960\,332\,040.562(26)           &   1152842742.711(17) \\
  26 &  955\,470\,615.084(34) &    4\,861\,425.4377(13) &  960\,332\,040.522(34)           &   1152842742.670(28) \\
  28 &  956\,140\,022.429(21) &    4\,192\,018.1260(12) &  960\,332\,040.555(21)           &   1152842742.704(06) \\
  30 &  956\,680\,116.330(34) &    3\,651\,924.1759(9) &  960\,332\,040.506(34)            &   1152842742.655(27) \\
  32 &  957\,122\,179.519(22) &    3\,209\,861.0067(8) &  960\,332\,040.526(23)            &   1152842742.674(22) \\ 
  35 &  957\,648\,684.552(25) &    2\,683\,355.9697(1) &  960\,332\,040.522(26)            &   1152842742.670(25) \\ 
  40 &  958\,277\,418.204(42) &    2\,054\,622.31805(33) &  960\,332\,040.522(42)          &   1152842742.671(42) \\ 
  45 &  958\,708\,525.814(21)$^{\rm b}$ &    1\,623\,514.66157(23) &  960\,332\,040.476(21)&   1152842742.624(21) \\ 
  50 &  959\,016\,922.718(35)$^{\rm b}$ &    1\,315\,117.77039(17) &  960\,332\,040.488(35)&   1152842742.637(35) \\ 
  53 &  959\,161\,558.215(16)$^{\rm b}$ &    1\,170\,482.28290(14) &  960\,332\,040.498(16)&   1152842742.647(16) \\ 
  55 &  959\,245\,118.319(24)$^{\rm b}$ &    1\,086\,922.13856(13) &  960\,332\,040.458(24)&   1152842742.606(24) \\ 
  60 &  959\,418\,690.453(19)$^{\rm b}$ &       913\,350.09968(10) &  960\,332\,040.553(19)&   1152842742.701(19) \\ 
  65 &  959\,553\,777.176(21)$^{\rm b}$ &       778\,263.35237(8) &  960\,332\,040.528(21) &   1152842742.677(21) \\ 
  70 &  959\,660\,968.990(51)$^{\rm b}$ &       671\,071.50269(6) &  960\,332\,040.493(51) &   1152842742.641(51) \\ 
  75 &  959\,747\,449.162(41)$^{\rm b}$ &       584\,591.35574(5) &  960\,332\,040.518(41) &   1152842742.666(41) \\ 
  80 &  959\,818\,229.110(37)$^{\rm b}$ &       513\,811.38941(4) &  960\,332\,040.499(37) &   1152842742.648(37) \\ 
  85 &  959\,876\,891.522(72)$^{\rm b}$ &       455\,149.00853(4) &  960\,332\,040.531(72) &   1152842742.679(72) \\ 
  90 &  959\,926\,052.374(57)$^{\rm b}$ &       405\,988.11880(3) &  960\,332\,040.493(57) &   1152842742.642(57) \\ 
  96 &  959\,975\,209.033(76)$^{\rm b}$ &       356\,831.51059(2) &  960\,332\,040.544(76) &   1152842742.692(76) \\ 
 102 &  960\,015\,949.716(114)$^{\rm b}$&       316\,090.74597(2) &  960\,332\,040.462(114)&   1152842742.611(114)\\ 
\hbox to 0pt{Average (weighted)}&&&                   960\,332\,040.532(21)                &   1152842742.680(021)\\             
\hbox to 0pt{QDT Extrap.}&&&                          960\,332\,040.501(32)\\           
\hbox to 0pt{Theory \cite{Pachucki2017}}&&&              960\,332\,038.0(2.0)\\           
\hbox to 0pt{Difference}&&&                  \phantom{960\,332\,03}2.5(2.0)\\
  \hline
  \hline
  \end{tabular}\\
  \hskip28pt\vbox{
  {\normalsize$^{\rm a}\,$Clausen et.al.\ \cite{Clausen2021}.}\\
  {\normalsize$^{\rm b}\,$Includes a Stark shift correction \cite{Clausen_Diss,ClausenMerkt2026}.}}
  \end{table*}
\begin{table*}[tb]  
\caption
{ Ionization energy of the $1s2s\;^3S_1$ state of $^4$He (in MHz).  
 $I_{\rm tot}(2\;^3S_1) = \nu_{\rm exp}(2\;^3S_1-n\;^3P_{\rm c}) + I_{\rm theo}(n\;^3P_{\rm c})$.} 
\label{table:results_triplet}
 \begin{tabular}{l r@{}l r@{}l r@{}l}
\hline
\hline
\multicolumn{1}{l}{$n$}& \multicolumn{2}{c}{$\nu_{\rm exp}(2\;^3S_0-n\;^3P_{\rm c})^{\rm a}$}  &
\multicolumn{2}{c}{$I_{\rm theo}(n\;^3P_{\rm c})$ } &\multicolumn{2}{c}{$I_{\rm tot}(2\;^3S_0)$}\\
\hline
  27&  1148\,307\,621&.274(11) &    4535\,121&.4498(10) &  1152\,842\,742&.723(12)\\
  29&  1148\,912\,959&.549(17) &    3929\,783&.1919(10) &  1152\,842\,742&.740(18)\\
  33&  1149\,809\,632&.367(13) &    3033\,110&.3603(9) &  1152\,842\,742&.727(14)\\
  35&  1150\,147\,008&.567(08) &    2695\,734&.1531(9) &  1152\,842\,742&.720(09)\\
  40&  1150\,779\,829&.987(07) &    2062\,912&.7491(4) &  1152\,842\,742&.736(08)\\
  50&  1151\,523\,381&.730(10) &    1319\,360&.9898(2) &  1152\,842\,742&.720(11)\\
  55&  1151\,752\,633&.012(14) &    1090\,109&.7147(1) &  1152\,842\,742&.727(15)\\
\hbox to 0pt{Average (weighted)}&&&&&                    1152\,842\,742&.727(11)\\[5pt]
\hbox to 0pt{$I(2\;^1S_0) + \nu(2\;^1S_0 - 2\;^3S_1)$}&&&&&   1152\,842\,742&.680(21)\\[5pt]
\hbox to 0pt{Grand average (weighted)}                &&&&&  1152\,842\,742&.705(16)\\ 
\hbox to 0pt{QDT Extrap.}&&&&&                           1152\,842\,742&.708(6)$^{\rm\, a}$\\[5pt]
\hbox to 0pt{$2\;^3S_1-8\;^3D_1$}&&&&&                   1152\,842\,742&.652(54)$^{\rm\, b}$\\
\hbox to 0pt{Theory \cite{Patkos2021}}&&&&&               1152\,842\,742&.231(52)\\
\hbox to 0pt{Difference}&&&&&                 \phantom{1152\,842\,74}0&.474(52)\\
  \hline
  \hline
  \end{tabular}
  \hskip45pt\vbox{
  $^{\rm a}$\normalsize Clausen et al.\ \cite{Clausen2025}.\\
  $^{\rm b}$\normalsize Wu and Wang \cite{Wu2025}.}
\end{table*}   
  
\section{Results}
The quantum defect extrapolations for the nonrelativistic energy can now be combined with the $1/n$ expansions for the relativistic and QED corrections to calculate absolute ionization energies for the high-lying $1snp\;^1P$ and $1snp\;^3P_c$ Rydberg states. This, combined with the measured transition frequencies from the low-lying $1s2s\;^1S_0$ and $1s2s\;^3S_1$ then yields ionization energies for the $S$-states according to
\begin{eqnarray}
I(2\;^1S_0) = I(n\;^1P_1)_{\rm theo} + h\nu(2\;^1S_0 - n\;^1P_1)\\
I(2\;^3S_1) = I(n\;^3P_c)_{\rm theo} + h\nu(2\;^3S_1 - n\;^3P_c)
\end{eqnarray}

Beginning with the singlet case, Table \ref{table:results_singlet}
lists total ionization energies obtained in this way for the 21 measured values of $n$ between 24 and 102. This extends the results obtained previously up to $n=35$ \cite{Drake2026}, and lowers the ionization energy slightly from  960\,332\,040.546(9) MHz to 960\,332\,040.532(21) MHz. For the triplet case, Table \ref{table:results_triplet} lists the similarly obtained total ionization energies for the measured 7 values of $n$ between 27 and 55.  This extends the results obtained previously up to $n=35$ \cite{Drake2026}, and raises the final result slightly from 1\,152\,842\,742.7247(78)stat(25)sys to 1152\,842\,742.727(11) MHz.  In addition, the table gives a value for $I(2\;^3S_1)$ obtained by adding the very accurately measured $2\;^1S_0 - 2\;^3S_1$ transition frequency \cite{Rengelink2018,Steinebach2026} 
to $I(2\;^1S_0)$ from Table \ref{table:results_singlet}.
The result of 1\,152\,842\,742.690(21) MHz lies $47\pm 34$ kHz lower, and so is statistically significant.  Finally, the grand average over both the 21 singlets and
7 triplets yields 1\,152\,842\,742.705(16) MHz.  This agrees extremely well with the corresponding value 1\,152\,842\,742.705(6) MHz obtained from a direct conventional QDT fit to the experimental data without adjustment, other than the use of $R_M^{(+)}$ for the Rydberg. Figure \ref{fig1} collects together all 28 measurements of the $2\;^3S_1$ ionization energy, and shows graphically the 9$\sigma$ discrepancy with theory \cite{Patkos2021} of $0.474 \pm0.052$ MHz.  The small but statistically significant difference between the singlet and triplet measurements is evident.  
\begin{figure}[tb]
\includegraphics[width=3.25in]{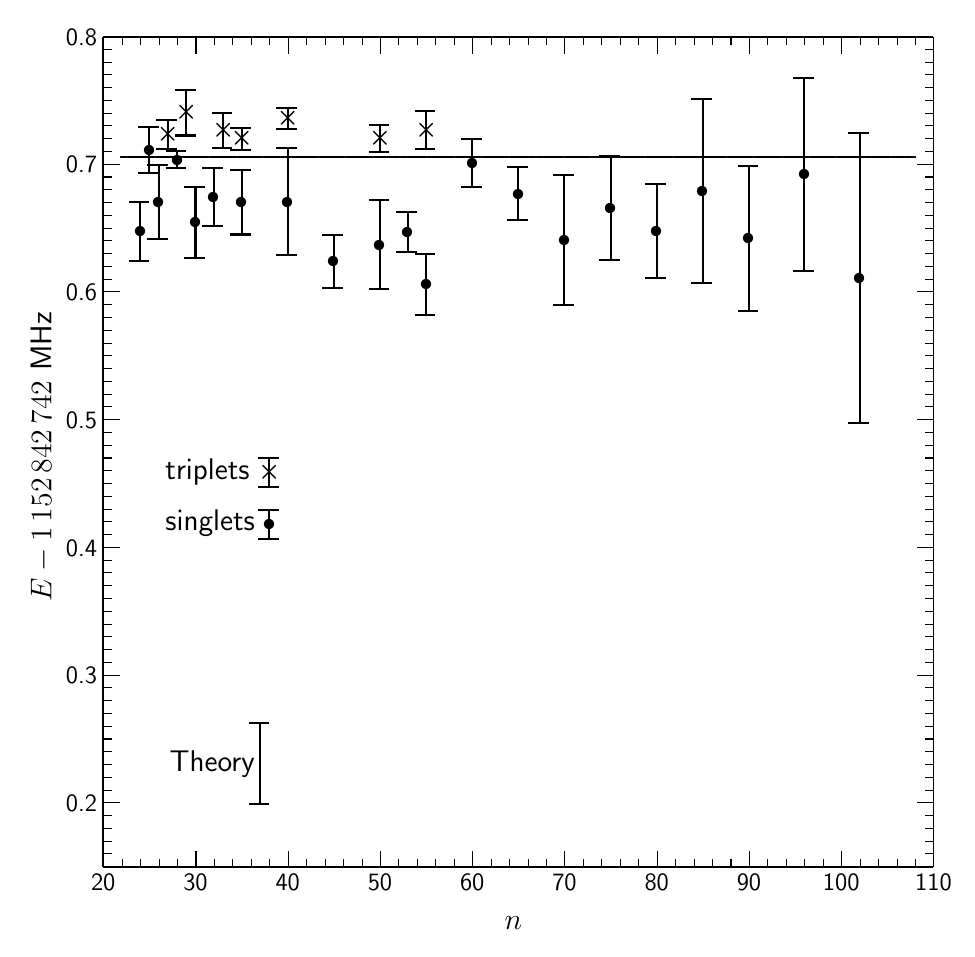}
\caption{Ionization energy of the $2\,^3S_1$ state of $^4$He as determined by adding the calculated ionization frequency of the $n\,^3P_{\rm c}$ state to the measured $2\,^3S_1-n\,^3P_{\rm c}$ transition frequency \cite{Clausen2025}, and similarly for the singlet case \cite{Clausen2021} together with the measured $2\,^1S_0-2\,^3S_1$ transition frequency 192\,510\,702.148\,649 MHz \cite{Rengelink2018,Steinebach2026}. The horizontal line represents the statistically weighted average.}
\label{fig1}
\end{figure} 

\section{Conclusions and Discussion}
The main approach if this work is to use a quantum defect fit only for the pure nonrelativistic binding energies, obtained by direct calculation up to $n=35$. The resulting quantum defect parameters shown in Table \ref{Table:QDfit} are of unprecedented accuracy, and reproduce the full 20-figure accuracy of the calculated energies over the entire range of $n$ up to 35.  The much smaller relativistic and QED corrections are represented to sufficient accuracy as purely $1/n$ expansions.  The expansions themselves reveal important contributions that vary as $1/n^2$ and come from second-order mass polarization terms of order $(\mu/M)^2$, either directly, or indirectly through recoil corrections to the terms $\tilde\Delta_2$ and $\langle B_1 + B_4\rangle$.  These $1/n^2$ terms were derived previously in the asymptotic limit of high-$L$ Rydberg states \cite{Drachman86,DrakeYan92}. The present work demonstrates that they are present also for $P$-states.  It remains to be shown that they are present with the same state-independent coefficients for $S$-states.  It is essential to subtract them before performing a quantum defect fit in order to avoid a $1/n^2$ contamination of the main $1/n*^2$ quantum defect fit.  These second-order recoil terms provide a formal justification for using the effective $R_M^{(+)}$ reduced mass Rydberg in place of the formally correct $R_M$ in performing quantum defect fits.    

There are two main points of significance to these results. First, each of the 28 measured transition frequencies \cite{Clausen2021,Clausen2025}
provides an independent measure of the $2\;^3S_1$ ionization energy, using a Ritz quantum defect fit only for the nonrelativistic energy for $n > 35$. The close agreement with the experimental data analyzed with a standard quantum defect fit \cite{Clausen2021,Clausen2025} provides a strong confirmation of QDT and the Ritz expansion, at least in the low-$Z$ region where relativistic effects are not too large.

The second significance is that the present results provide an unequivocal confirmation of a 9$\sigma$ disagreement with the theoretical ionization energy of the $2\;^3S_1$ state \cite{Patkos2021} of  $0.474 \pm0.052$ MHz, wth experiment lying lower than theory.  The discrepancy is particularly puzzling since it is almost the same for the case of $^3$He \cite{Clausen2025c}, and it 
is not apparent in the $2\;^3S_1-2\;^1S_0$ isotope shift \cite{Eikema2025,Steinebach2026}. It therefore does not depend on the particular isotope or neutron number.  In the absence of errors on the theoretical side, or unexpectedly large higher-order QED terms, Cong et al.\ \cite{Cong2026} discuss possible explanations involving new particles affecting the electron-electron interaction. They are able to exclude two of the four possible interactions just based on the sign of the discrepancy for the $2\;^3S_1$ state, and most of the remainder, except possibly for a light scalar boson as the only viable candidate. It is not yet clear whether the discrepancy is also present for the singlet spectrum since the same $m\alpha^7$ terms 
\cite{Patkos2021} have not yet been calculated to reduce the theoretical uncertainty.  Parallel high precision experiments and theory for the triplet spectrum of Li$^+$ \cite{Yerokhin2023} would be very interesting as the next case to study.

{\it Note Added:}  After this work was completed, we learned of calculations based on the correlated B-spline method \cite{Chi2026} that are in similarly good agreement with the experimental QD extrapolation of Clausen et al. Close agreement with the present calculation provides an important confirmation of both sets of results.    

\begin{center}
\bf DATA AVAILABILITY
\end{center}
The data supporting the findings of this article have
been tabulated within the article, with additional tabuations of contributions to the total ionization energy 
for each $n$ up to 35 in Ref.\ \cite{data}. 
Additional metadata
are available from the corresponding author upon request.

\begin{acknowledgements}
This research was supported by the Natural Sciences and Engineering Research Council of Canada (NSERC) and by the Digital Research Alliance of Canada/Compute Ontario. The authors are grateful to Gloria Clausen Fr\'ed\'eric Merkt for providing the Stark-shift corrections for the measured ransition frequencies n Table XI. 
\end{acknowledgements}


\newpage
\begin{references} 
\bibitem{Clausen2021} 
G. Clausen, P. Jansen, S. Scheidegger, J. A. Agner, H. Schmutz, and F. 
Merkt, Ionization energy of the metastable $2\;^1S_0$ state of $^4$He 
from Rydberg-series extrapolation, Phys.\ Rev.\ Lett.\ {\bf 127}, 
093001 (2021). 
\bibitem{Clausen2025} 
G. Clausen, K. Gamlin, J. A. Agner, H. Schmutz, and F. Merkt, Metrology 
in a two-electron atom: The ionization energy of metastable triplet 
helium ($2\;^3S_1$), Phys.\ Rev.\ A {\bf 111}, 012817 (2025). 

\bibitem{Pachucki2017} 
K. Pachucki, V. Patk\'os, and V. A. Yerokhin, Testing fundamental 
interactions on the helium atom, Phys.\ Rev.\ A {\bf 95}, 062510 
(2017). 
\bibitem{Patkos2021} 
V. Patk\'o\v{s}, V.A. Yerokhin and K. Pachucki, Complete $m\alpha^7$ 
Lamb shift of helium triplet states, Phys.\ Rev.\ A {\bf 103}, 042809 
(2021). 
\bibitem{Yan1995} 
Z.-C. Yan and G. W. F. Drake, High Precision Calculation of Fine 
Structure Splittings in Helium and He-Like Ions, Phys.\ Rev.\ Lett.\ 
{\bf 74}, 4791 (1995). 
\bibitem{Pachucki2012} 
K. Pachucki, V. A. Yerokhin, and P. Cancio Pastor, Quantum 
electrodynamic calculation of the hyperfine structure of $^3$He, Phys.\ 
Rev.\ A {\bf 85}, 042517 (2012). 
\bibitem{Pachucki2010} 
K. Pachucki and V. A. Yerokhin, Fine Structure of Heliumlike Ions and 
Determination of the Fine Structure Constant, Phys.\ Rev.\ Lett.\ {\bf 
104}, 070403 (2010). 
\bibitem{Zheng2017} 
X. Zheng, Y. R. Sun, J.-J. Chen, W. Jiang, K. Pachucki, and S.-M. Hu, 
Laser Spectroscopy of the Fine-Structure Splitting in the $2\;^3P_J$ 
Levels of $^4$He, Phys.\ Rev.\ Lett.\ {\bf 118}, 063001 (2017). 
\bibitem{Kato2018} 
K. Kato, T. D. G. Skinner, and E. A. Hessels, Ultrahigh-Precision 
Measurement of the $n=2$ Triplet $P_J$ Fine Structure of Atomic Helium 
Using Frequency-Offset Separated Oscillatory Fields, Phys.\ Rev.\ 
Lett.\ {\bf 121}, 143002 (2018). 
\bibitem{Bondy2025} 
A. T. Bondy, G. W. F. Drake, C. Mcleod, E. M. R. Petrimoulx, X.-Q.  Qi, 
Z.-X. Zhong, Theory for the Rydberg states of helium: Comparison with 
experiment for the $1s24p\;^1P_1$ state $(n=24)$, Phys.\ Rev.\ A {\bf 
111}, L010803 (2025). 
\bibitem{Drake2026} 
G. W. F. Drake, A. T. Bondy, O. P. Hallett, and B. C. Najem, Theory for 
the Rydberg states of helium: Results for $2 \le n \le 35$ and 
comparison with experiments for the singlet and triplet P states, 
Phys.\ Rev.\ A {\bf 113}, 012810 (2026). 

\bibitem{Chi2025} 
J. Chi, H. Fang, Y.-H. Zhang, X.-Q. Qi, L.-Y. Tang and T.-Y. Shi, 
Accurate Nonrelativistic Energy Calculations for Helium $1snp\;^{1,3}P$ 
$(n = 2 to 27)$ States via Correlated B-Spline Basis Functions, Atoms 
{\bf 13}, 72 (2025). 

\bibitem{Fang2026} 
H. Fang, J. Chi,X.-Q. Qi, Y. H. Zhang, L.-Y. Tang, and T.-Y. Shi, 
Precise ab initio calculations of $^4$He ($1snp\;^3P_J$) fine structure of high 
Rydberg states Phys.\ Rev.\ A {\bf 113}, 012812 (2026). 

\bibitem{Rengelink2018} 
R. J. Rengelink, Y. van der Werf, R. P. M. J. W. Notermans, R. Jannin, 
K. S. E. Eikema, M. D. Hoogerland, and W. Vassen, Precision 
spectroscopy of helium in a magic wavelength optical dipole trap, Nat.\ 
Phys.\ {\bf 14}, 1132 (2018). 
\bibitem{Steinebach2026} 
K. Steinebach, J. C. J. Koelemeij, H. L. Bethlem, and K. S. E. Eikema, 
Spectroscopy of $^4$He at 0.25 ppt Uncertainty and Improved 
Alpha-Helion Charge-Radius Difference Determination, arXiv: 2601.19444 
(2026). 
\bibitem{DrakeAdv} 
G. W. F. Drake, ``Quantum defect theory and analysis of high precision 
helium term energies," in {\it Advances in Atomic, Molecular and 
Optical Physics} (Academic Press, 1993) Vol.\ 32, pp.\ 93--116. 

\bibitem{DrakeYan92} 
G. W. F. Drake and Z.-C. Yan, Energies and relativistic corrections for 
the Rydberg states of helium: Variational results and asymptotic 
analysis, Phys.\ Rev.\ A, {\bf 46}, 2378 (1992). 
\bibitem{Drake87} 
G. W. F. Drake, New variational techniques for the 1snd states of 
helium, Phys.\ Rev.\ Lett.\ {\bf 59}, 1549 (1987). 

\bibitem{DrakeMakowski88} 
G. W. F. Drake and A. J. Makowski, High Precision eigenvalues for the 
$1s2p\ ^1P$ and $^3P$ states of helium, J. Opt.\ Soc.\ Am.\ B, {\bf 5}, 
2207 (1988). 
\bibitem{CODATA} 
P. J. Mohr, D. B. Newell, B. N. Taylor, and E. Tiesinga, Codata 
recommended values of the fundamental physical constants: 2022, Rev.\ 
Mod.\ Phys.\ {\bf 97}, 025002 (2025). 
\bibitem{Bailey}
DQFUN: A double-quad precision package with special functions,
Available from https://www.davidhbailey.com/dhbsoftware/README-dqfun.txt [accessed: 2010-09-30].
\bibitem{Bethe} 
H. A. Bethe and E. E. Salpeter, {\it Quantum Mechanics of One- and 
Two-Electron Atoms}, (Springer, New York, 1957), p.\ 181. 
\bibitem{Stone1963} 
A. P. Stone, Proc. Phys. Soc. {\bf 77}, 786 (1961); {\bf 81}, 868 
(1963). 
\bibitem{Kabir} 
P. K. Kabir and E. E. Salpeter, Radiative Corrections to the 
Ground-State Energy of the Helium Atom, Phys.\ Rev.\ {\bf 108}, 1256 
(1957). 
\bibitem{Araki} 
H. Araki, Quantum-Electrodynamical Corrections to Energy-Levels of 
Helium, Prog.\ Theo.\ Phys.\ {\bf 17}, 619 (1957). 

\bibitem{Sucher} 
J. Sucher, Energy Levels of the Two-Electron Atom to Order $\alpha^3$ 
Ry; Ionization Energy of Helium Phys.\ Rev.\ {\bf 109}, 1010 (1958). 
\bibitem{YanDrake2003} 
Z.-C. Yan and G. W. F. Drake, Bethe Logarithm and QED Shift for 
Lithium, Phys.\ Rev.\ Lett.\ {\bf 91}, 113004 (2003). 

\bibitem{PachuckiKomasa2004} 
K. Pachucki and J. Komasa, Relativistic and QED Corrections for the 
Beryllium Atom, Phys.\ Rev. Lett.\ {\bf 92}, 213001 (2004). 
\bibitem{Korobov2019} 
V. I. Korobov, Bethe logarithm for the helium atom, Phys.\ Rev.\ A {\bf 
100}, 012517 (2019). 

\bibitem{Lesiuk2024} 
M. Lesiuk and J. Lang, Atomic Bethe logarithm in the mean-field 
approximation, Phys.\ Rev.\ A {\bf 108}, 042817 (2023).  The values 
obtained for $\ln( k_0/Z^2)$ for atoms heavier than Be are larger than 
the hydrogenic value by up to 4.7\%. 
\bibitem{Drake2001} 
G. W. F. Drake, QED Effects in Helium and Comparisons with High 
Precision Experiment, Phys.\ Scr.\ {\bf T95}, 22 (2001). 

\bibitem{Pachucki2024} 
K. Pachucki, V. Lensky, F. Hagelstein, S. S. L. Muli, S. Bacca, and R. 
Pohl, Comprehensive theory of the Lamb shift in light muonic atoms 
Rev.\ Mod.\ Phys. {\bf 96}, 015001 (2024). 


\bibitem{data} 
Further tables of matrix elements for all $n\,P$ states of helium with 
$2\le n \le 35$ are available at DOI: 10.5061/dryad.cjsxksnkv. 

\bibitem{Edlen} 
B. Edl\'en, Atomic spectra, in {\it Encyclopedia of Physics}, {\bf 27}, 
80 (Springer-Verlag, Berlin, 1964). 
\bibitem{Seaton} 
M. J. Seaton, Quantum Defect Theory, Rep.\ Prog.\ Phys.\ {\bf 46}, 167 
(1983). 
\bibitem{Jungen} 
C. Jungen, Elements of Quantum Defect Theory, in {\it Handbook of 
High-resolution Spectroscopy}, Edited by M. Quack and F. Merkt, (John 
Wiley \& Sons, Ltd., Chichester, UK., 2011). 
\bibitem{Hartree} 
D. R. Hartree, The Wave Mechanics of an Atom with a non-Coulomb Central 
Field. Part III. Term Values and Intensities in Series in Optical 
Spectra, Cambridge Philos.\ Soc., {\bf 24}, 426 (1028). 
\bibitem{Langer} 
R. M. Langer, A Generalization of the Rydberg Formula, Phys.\ Rev.\ 
{\bf 35}, 649 (1930). 
\bibitem{Advances} 
G. W. F. Drake, Quantum Defect Theory and Analysis of High-Precision 
Helium Term Energies, Adv.\ At.\ Mol.\ Opt.\ Phys.\ {\bf 32}, 93 
(1994). 
\bibitem{Seaton1966} 
M. J. Seaton, Quantum defect theory II: Illustrative one-channel and 
two-channel problems, Proc.\ Phys.\ Soc.\ {\bf 88}, 815 (1966). 
\bibitem{SwainsonDrake} 
G. W. F. Drake and R. A. Swainson, Quantum defects and the $1/n$ 
dependence of Rydberg energies: Second-order polarization effects, 
Phys.\ Rev.\ A {\bf 44}, 5448 (1991). 
\bibitem{Johnson77} 
W. R. Johnson and K. T. Cheng, Quantum defects for highly stripped 
ions, J. Phys.\ B: At.\ Mol.\ Phys.\ {\bf 12}, 863 (1979). 
\bibitem{Jacobs2022} 
D. M. Jacobs, Relativistic Ritz approach to hydrogenlike atoms: 
Theoretical considerations, Phys.\ Rev.\ A {\bf 106}, 062810 (2022). 
\bibitem{Drachman86} 
R. J. Drachman, Rydberg states of helium: Nuclear-recoil corrections, 
Phys.\ Rev.\ A {\bf 33}, 2780 (1986). 
\bibitem{Clausen_Diss} 
DISS. ETH NO. 31047 G. Clausen, Precision spectroscopy in $^3$He and 
$^4$He: Rydberg states and ionization energies, Ph.D. Dissertation (ETH 
Zurich No.\ 31047, 2025). 
\bibitem{ClausenMerkt2026} 
G. Clausen and F. Merkt, private communication. 
\bibitem{Wu2025} 
M.-H. Wu and L.-B. Wang, Frequency measurement of the $2\;^3S_1 
8\;^3D_1$ two-photon transition in atomic $^4$He, Phys.\ Rev.\ A {\bf 
111}, 052809 (2025). 

\bibitem{Clausen2025c} 
G. Clausen and F. Merkt, Ionization Energy of Metastable $^3$He 
($2\,^3S_1$) and the Alpha- and Helion-Particle Charge-Radius 
Difference from Precision Spectroscopy of the $np$ Rydberg Series, 
Phys.\ Rev.\ Lett.\ {\bf 134}, 223001 (2025). 

\bibitem{Eikema2025} 
Y. van der Werf, K. Steinebach, R. Jannin, H. L. Bethlem, and K. S. E. 
Eikema, Alpha and helion particle charge radius difference determined 
from quantum-degenerate helium, Science, {\bf 388}, 850, 854 (2025). 
\bibitem{Cong2026} 
L. Cong, F. Ficek, R. Abdullin, M. G. Kozlov, and D. Budker, Testing 
Exotic Electron Electron Interactions with the Helium Ionization-Energy 
Anomaly, Phys.\ Rev.\ A, submitted (1926). 
\bibitem{Yerokhin2023} 
V. A. Yerokhin, V. Patk\'o\v{s}, and K. Pachucki, QED $m\alpha^7$ 
effects for triplet states of heliumlike ions, Phys.\ Rev.\ A {\bf 
107}, 012810 (2023). 
\bibitem{Chi2026} 
J. Chi, H. Fang, Y.-H. Zhang L.-Y. Tang, and T.-Y. Shi, Ionization 
energies for Rydberg $^4$He $(1snp\;^{1,3}P)$ states using the 
correlated B-spline basis function method, Phys.\ Rev.\ A. submitted. 
\end{references}
\end{document}